\documentclass[a4paper,12pt,twoside]{article} 
\usepackage[total={11in,8.5in}]{geometry}

 \usepackage[authoryear,round]{natbib}
\usepackage{multirow}
\usepackage{algorithm}
\usepackage[noend]{algpseudocode}
\usepackage{setspace}
\usepackage{lipsum}

 \usepackage[usenames]{color}
\usepackage[usenames,dvipsnames]{xcolor}
 \usepackage{textcomp}
\usepackage[latin1]{inputenc}  

\usepackage{color,soul}
\usepackage{float}
\usepackage{fancyvrb}
\usepackage{hyperref}
 \usepackage{verbatim}
\hypersetup{linkbordercolor=1 1
1,colorlinks=false,linkcolor=blue,pdfborder=0 0 0}
\usepackage{bbold}
\usepackage{xcolor}
\usepackage{lipsum}

\usepackage[catalan,spanish,english]{babel}
\usepackage{lscape}
\usepackage{latexsym, graphicx, graphics}
\usepackage{amsmath}
\usepackage{mathtools}
\usepackage[usenames]{color}	
\usepackage{tikz}
\usepackage{amssymb}
\usepackage[labelfont=bf,labelsep=period,justification=raggedright]{caption}
\usepackage{tabularx}
\usepackage{subfig}
\usepackage{graphicx}
\makeatletter
\def\BState{\State\hskip-\ALG@thistlm}
\makeatother
\date{}

\renewenvironment{abstract}
 {\footnotesize
  \begin{center}
  \bfseries \abstractname\vspace{-.7em}\vspace{0pt}
  \end{center}
  \list{}{
    \setlength{\leftmargin}{.7cm}%
    \setlength{\rightmargin}{\leftmargin}%
  }%
  \item\relax}
 {\endlist}

\begin{document}

\graphicspath{{./GRAPHICS/}}

\pagenumbering{arabic}
\selectlanguage{english}

\pagestyle{plain}
{\center {\Large
\textbf{Joint Estimation of Sparse Networks with\\
application to Paired Gene Expression data}\\
}}
\vspace{0.2cm}
\hspace{-0.6cm}
Adria Caballe Mestres$^{a,b}$, Natalia Bochkina$^a$ and Claus Mayer$^b$\\
$^a$ University of Edinburgh \& Maxwell Institute, Scotland, UK \\
$^b$ Biomathematics \& Statistics Scotland, Scotland, UK

	\doublespacing
	\begin{abstract}
{\footnotesize We consider a method to jointly estimate sparse precision matrices and their underlying graph structures using dependent high-dimensional datasets. We present a penalized maximum likelihood estimator which encourages both sparsity and similarity in the estimated precision matrices where tuning parameters are automatically selected by  controlling the expected number of false positive edges. We also incorporate an extra step to remove edges which represent an overestimation of triangular motifs. We conduct a simulation study to show that the proposed methodology presents consistent results for different combinations of sample size and dimension. Then, we apply the suggested approaches to a high-dimensional real case study of gene expression data with samples in two medical conditions, healthy and colon cancer tissues, to estimate a common network of genes as well as the differentially connected genes that are important to the disease. We find denser graph structures for healthy samples than for tumor samples, with groups of genes interacting together in the shape of clusters. 
}

	\end{abstract}
\doublespacing

\vspace{-0.7cm}
\section{Motivating problem in genomic data}%
Genomic data produced by high-throughput technology are nowadays easy to collect and store 
generating many statistical questions. For instance here we want to use a dataset where 
genomic profiles are obtained for individuals in different classes.  It is publicly available in the ArrayExpress 
database (\url{http://www.ebi.ac.uk/arrayexpress/}) and it is formally presented in \cite{Hinoue2012}. 
It contains the gene expression information of 25 patients in two samples (tissues) for each 
gene/patient: the expression in a colon cancerous tissue and the expression in its adjacent 
healthy tissue. In total, there are more than 24k genes. 

One of the challenges in the analysis of this data is to understand how genes interact between 
each other in a cell as well as detecting which groups of such connections vary from a healthy 
to a cancer state. This can be formulated by an estimation problem of sparse conditional dependence 
networks which are fully characterized by their underlying precision matrices (inverse of covariance/correlation 
matrices). Two genes are said to be conditionally independent given all the remaining genes 
if their correspondent coefficient in the precision matrix is zero. 

This type of estimation problem is extensively studied in the field of statistics and also bioinformatics  
when the data is high-dimensional (dimension is larger than the sample size) in which case 
maximum likelihood estimators are not suitable \citep{Pourahmadi2007}. Methods that address this issue 
to estimate a single precision matrix include sparsity-penalization approaches known as graphical lasso 
\citep{Meinshausen2006, Friedman2007, Peng2009}.
A natural extension is applied to jointly estimate multiple precision matrices by using an additional
penalization term that encourages the similarity between such matrices.
For instance, \cite{Guo2011} use a group-lasso penalization (GGL) or \cite{Danaher2014}  incorporate 
a fused-lasso penalization option (FGL). The FGL method, which we will consider as 
the current method, yields better graph recovery rates than estimating the matrices separately 
when these are expected to be similar. However, it is designed under the assumption of independence 
between datasets.

Motivated by real data, in this paper we extend the methodology presented in  \cite{Danaher2014} for a 
more general case of estimating jointly two sparse-similar precision matrices whose datasets are dependent. 
A related proposal is given in \cite{Wit2015} who estimate a joint precision matrix that can reflect 
several time points in a disease process. The divergence with respect to our method is in the interpretation of the 
differential network. Our aim is to find out which gene associations are (not) common between populations 
whereas they consider such matrix design as a constrain prior to estimation. 

In Section \ref{SEC2} we describe the estimation problem, in Section \ref{tuning} we propose a 
strategy to select the tuning parameters originated by the penalization terms and in Section 
\ref{triangle1}  we discuss an issue in the current algorithm to estimate triangular motifs structures. 
In Section \ref{SEC4} we apply all the methodology to simulated datasets given different models to 
generate the data. Finally, in Section \ref{SEC5} we estimate conditional dependence structures for the 
motivating application to colon cancer gene expression data.  The proposed methodology is implemented within the R package {\tt ldstatsHD} \citep{Caballe2016b}.

\section{Weighted fused graphical lasso}\label{SEC2}
\subsection{Problem set up and cross-correlation}
We assume that the data are independent and identically distributed (iid) observations 
from a Gaussian model, $ [X_k,Y_k] \sim N_{2p}(0,\Omega^{-1})$, $k=1,\ldots, n$, with dimension $p$ and sample 
size $n$, assuming, without a loss of generality, that the mean is zero.  The matrix $\Omega$ represents the joint 
conditional dependence structure for $X$ and $Y$, and it is defined by
\begin{equation}\label{eq:jointCor}
\Omega = \Sigma^{-1} = \begin{bmatrix} \Sigma_X & \Sigma_{XY} \\ \Sigma_{XY}^t  & \Sigma_Y \end{bmatrix}^{-1} = \begin{bmatrix} \Omega_X^J & \Omega_{XY}^J \\ \Omega_{YX}^J  & \Omega_Y^J \end{bmatrix}.
\end{equation}
The objective is to estimate the precision matrices $\Omega_X= \Sigma_X^{-1}$, $\Omega_Y= \Sigma_Y^{-1}$ and $\hat{\Omega}_d = \hat{\Omega}_Y-\hat{\Omega}_X$ with the assumptions of sparsity and similarity among the two matrices.

If $\Sigma_{XY} = 0$ (which means that $\Omega_{XY}^J$ must  also be $0$), then pairs $X$ and $Y$ are 
independent with both $\Omega_X= \Omega_X^J$ and $\Omega_Y= \Omega_Y^J$. 
In contrast, if $\Sigma_{XY}$ contains at least one non-zero element, then $X$  and $Y$ are 
dependent (e.g. in paired data) and the equality does not hold. 

\begin{figure}[ht]
\begin{center}
\includegraphics[width=5cm,height=5cm]{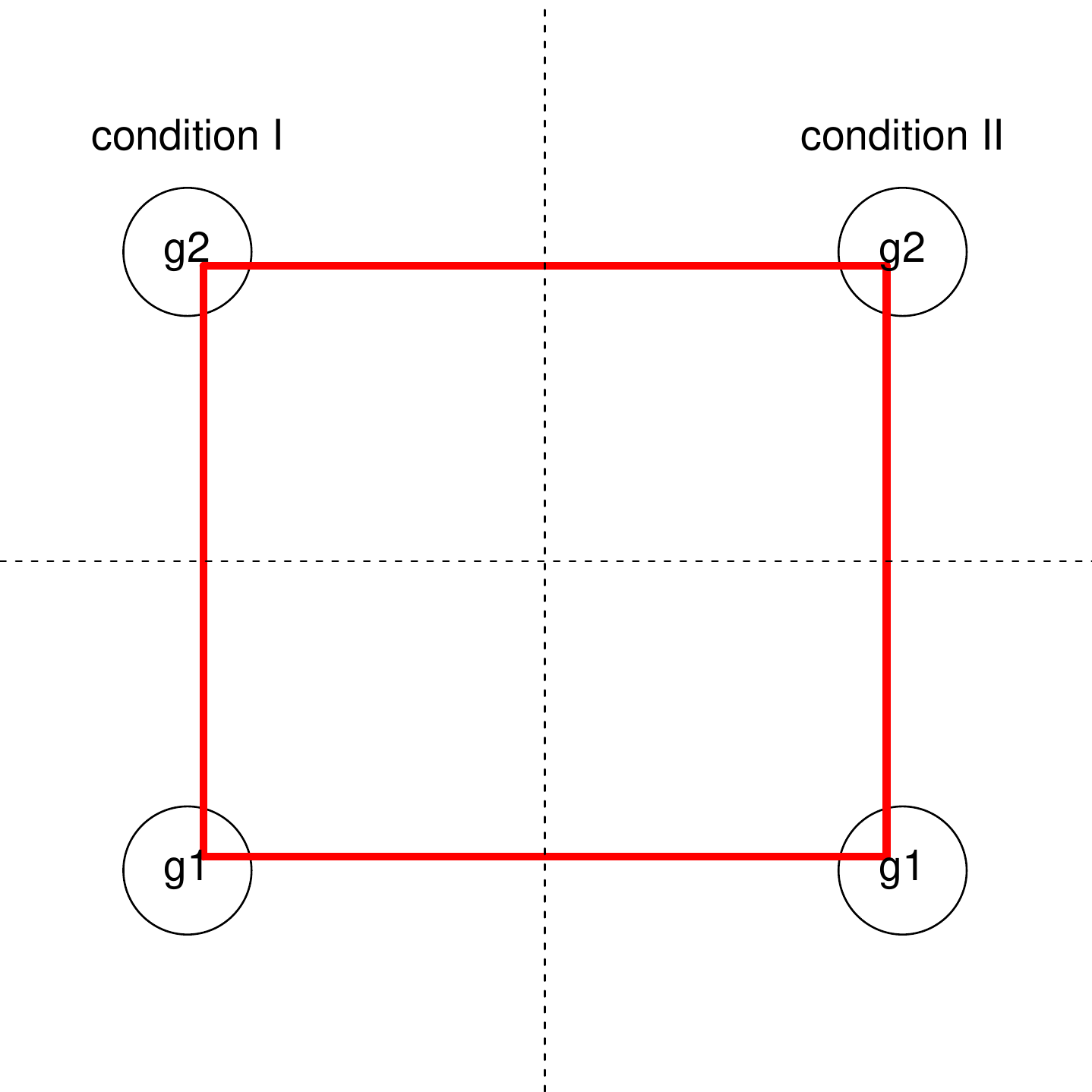} \caption{\footnotesize{Square-type conditional graph dependence structure. Conditional independence between different genes relating the two conditions.}}\label{fDiagOmega}
\end{center}
\end{figure}

There are various types of dependence structures defined in either $\Sigma_{XY}$ or $\Omega_{XY}$.  For instance, structures defined in the cross-covariance matrix $\Sigma_{XY}$ motivated by an additive model ($\Sigma_{XY} = \Delta \Sigma_X \Delta^t$, with diagonal matrix $\Delta$, $0\leq\Delta_{ii}<1$) or a multiplicative model  ($\Sigma_{XY} = \Delta \Sigma_X^{1/2}\Sigma_Y^{1/2} \Delta^t$).  Both such cases, which coincide when $\Sigma_X=\Sigma_Y$,  can be described by a Kronecker products formulation \citep{Fan2008, Srivastava2008}. \cite{Wit2015} make  a further simplification in the dependence structure by characterizing the cross-precision matrix  $\Omega_{XY}^J$. They assume that $\Omega_{XY}^J$ is a diagonal matrix. Hence, that any variable of the first dataset $X_{ki}$ is independent from any variable of the other dataset $Y_{kj}$ if $i\neq j$, $k \in 1,\ldots,n$, once conditioning on variables $X_{kj}$ and $Y_{ki}$ (i.e. see Figure \ref{fDiagOmega}  for the underlying graphical representation). The main results we give in the next subsections can be applied for any type of dependence structure. However, for the real data analysis and also simulations we assume that the cross partial correlation matrix is diagonal.

\vspace{-0.8cm}
\subsection{Weighted fused graphical lasso for two dependent datasets }\label{GM}
We propose to use a weighted-fused graphical lasso (WFGL) maximum likelihood estimator for the joint precision matrix: 
\begin{equation}\label{DAN}
\hat{\Omega}_{WFGL}^{\lambda} = \arg\max\limits_{\Omega_X,\Omega_Y} [\sum_{m = X,Y} \log\det\Omega_m -tr(\Omega_m S_m) - P_{\lambda_1,\lambda_2, V}(\Omega_X, \Omega_Y)],
\end{equation}
with 
\begin{equation}\label{DANpen2}
P_{\lambda_1,\lambda_2, V}(\Omega_X,\Omega_Y) = \lambda_1||\Omega_X||_1 + \lambda_1||\Omega_Y||_1 +\lambda_2\sum_{i=1}^p\sum_{j=1}^p v_{ij} |\Omega_{Y_{ij}}-\Omega_{X_{ij}}|,
\end{equation}
where $\lambda_1$ is the sparsity tuning parameter, $\lambda_2$  is the similarity tuning parameter, and $V=[v_{ij}]$ is a $p\times p$ matrix to weight $\lambda_2$ for each coefficient of the differential precision matrix. In case $v_{ij}=1$ for all pairs $(i,j)$, then the maximization problem coincides with  the fused graphical lasso (FGL) presented in \cite{Danaher2014}. As novelty, in the next subsection we define weights that account for the dependence structure between the two datasets. 

The maximization problem in  \eqref{DAN} and \eqref{DANpen2} can be solved by the ADMM-type algorithm \citep{Boyd2010}  described in Algorithm \ref{algJGL}. The main difference with respect to the FGL algorithm is in step 6, where different similarity penalties are considered to estimate the differential network. 

\vspace{0.5cm}
\begin{algorithm}[ht]
\caption{Weighted Fused Graphical Lasso}\label{algJGL}
\begin{algorithmic}[1]
 \Procedure{WFGL}{$\lambda_1,\lambda_2, \rho, V$}
\BState Define the Lagrangian formulation of the maximization problem in (\ref{DAN}):
\begin{equation}
L_{\rho} =  -[\sum_{m=X,Y} \log\det\Omega_m -tr(\Omega_m S_m) + P_{\lambda_1,\lambda_2,V}(A_X,A_Y) + 
 \frac{\rho}{2}\sum_{m=X,Y} ||\Omega_m - A_m +U_m||^2_F],
\end{equation}
where $U_m$ are dual variables, $A_m$ corresponds to $\Omega_m$ and $\rho$ is a positive constant that is used as a regularization parameter with default value equal to 1. 

\BState Initialization: set iteration $t=0$, $U_m^{(t)} = 0$ and $\hat{S}_m^{(t)} = S_m$ corresponding to the sample covariance matrix for $m=X,Y$. Repeat 4-8 until convergence.
\BState Find $\hat{\Omega}_m^{(t)}$ using a quadratic regularized inverse as shown in \cite{Witten2009}. Given the eigenvalue decomposition of matrix  $\hat{S}_m^{(t)} = V_m^{(t)}D_m^{(t)}V_m^{'(t)}$, the inverse is found by
\begin{equation}\label{regInv}
 \hat{\Omega}_m^{(t)}=V_m^{(t)}\tilde{D}_m^{(t)}V_m^{'(t)}, \mbox{\hspace{0.5cm}} \tilde{D}_{m_{jj}}^{(t)} = \frac{n}{2\rho}\left(-D_{m_{jj}}^{(t)} + \sqrt{(D_{m_{jj}}^{(t)})^2 + 4\rho /n}\right),
\end{equation}

\BState Find $[\hat{A}_X^{(t)},\hat{A}_Y^{(t)}]$ by minimizing $
\frac{\rho}{2}\sum_{m=X,Y}||A_m-(\hat{\Omega}^{(t)}_m+U^{(t)}_m)||^F_2 + P_{\lambda_1,\lambda_2,V}(A_X,A_Y)
$ using a thresholding approach:

\State Given $\{\hat{A}_m^{'(t)} = \hat{\Omega}^{(t)}_m+U_m^{(t)}\}_{m=X,Y}$, set equal precision matrix elements if the 

\hspace{-0.3cm} estimated differences are smaller than $V \lambda_2/\rho$:
\begin{equation}\label{simEq}
[\hat{A}_{X_{ij}}^{''(t)},\hat{A}_{Y_{ij}}^{''(t)}] =  \left\{ \begin{array}{r l l}
      &   [.5(\hat{A}_{X_{ij}}^{'(t)} + \hat{A}_{Y_{ij}}^{'(t)}),.5(\hat{A}_{X_{ij}}^{'(t)} + \hat{A}_{Y_{ij}}^{'(t)})]& \mbox{if $v_{ij}|\hat{A}_{X_{ij}}^{'(t)} -\hat{A}_{Y_{ij}}^{'(t)}|\leq\lambda_2/\rho$};\\
      &   [\hat{A}_{X_{ij}}^{'(t)} + v_{ij}\lambda_2/\rho,\hat{A}_{Y_{ij}}^{'(t)} - v_{ij}\lambda_2/\rho]    &
\mbox{if $\hat{A}_{X_{ij}}^{'(t)} - \hat{A}_{Y_{ij}}^{'(t)}> v_{ij}\lambda_2/\rho$};\\
      &   [\hat{A}_{X_{ij}}^{'(t)}  - v_{ij}\lambda_2/\rho,\hat{A}_{Y_{ij}}^{'(t)} + v_{ij}\lambda_2/\rho]    &
\mbox{if $\hat{A}_{X_{ij}}^{'(t)} -\hat{A}_{Y_{ij}}^{'(t)}< -v_{ij}\lambda_2/\rho$};\\
         \end{array} \right.
\end{equation}
\State Set precision matrix elements to zero by soft-thresholding \citep{Rothman2009} 

\hspace{-0.3cm} with threshold given by $\lambda_1$:
\begin{equation}\label{spaEq}
\hat{A}_{m_{ij}}^{(t)}  = \text{sign}(\hat{A}_{m_{ij}}^{''(t)}) \left( |\hat{A}_{m_{ij}}^{''(t)}| - \lambda_1  \right)_{+}, \mbox{\hspace{0.5cm}}  m=X,Y.
\end{equation}

\BState  Set $t = t +1$. Update $U^{(t)}_m = U^{(t-1)}_m + (\hat{\Omega}^{(t-1)}_m - \hat{A}^{(t-1)}_m)$ and $\hat{S}^{(t)}_m = S_m - \frac{\rho}{n} \hat{A}^{(t-1)}_m  + \frac{\rho}{n} U^{(t)}_m$ for $m=X,Y$.

\EndProcedure
\end{algorithmic}
\end{algorithm}

\subsection{Weights in the similarity penalization term}\label{pairedWeights}
We consider the weights as a way to marginally normalize the initial estimated differential precision matrix which can be found in step 6 of Algorithm \ref{algJGL} just before thresholding.
Our objective is to adapt the similarity-penalty parameter for each pair $(i,j)$ such that the probability to recover differential edges is independent of the relationship between variables in the two datasets.

We define the partial correlation matrix $\hat{W} = [\hat{w}_{ij}]$ (scaled of estimated precision matrix $\hat{\Omega}$) and the Fisher transformation function $g: \mathbb{R}  \to \mathbb{R}$, $g(z) = \log\{(1+z)/(1-z)\}/2$.
We propose to use weights $V = [v_{ij}]$ described by
\begin{equation}
v_{ij} = \text{var}[ g(\hat{w}_{Y_{ij}}) - g(\hat{w}_{X_{ij}})]^{-1/2} \doteq \text{const} (2 - 2\psi_{ij})^{-1/2},
\end{equation}
where  $\psi_{ij} = \text{cor}( g(\hat{w}_{X_{ij}}), g(\hat{w}_{Y_{ij}}))$. 
Individual variances expressed by $\text{var}( \hat{\Omega}_{m_{ij}})$, for both $m=X,Y$, could also be included in the weights. However,  it can be proved that elements with large variances are, generally, more likely to contain non-zero 
coefficients than  elements with low variances. Hence, correcting in this case might increase the number of  false positive edges.

The asymptotic expression for the correlation of Fisher transform sample correlation coefficients $\hat{R} = [\hat{r}_{ij}]$ (scaled of sample covariance matrix elements) is derived in \cite{Elston1975} and \cite{Olkin1990}, among others, and it is only function of the true correlation coefficients. Similarly, we derive the asymptotic expression for the correlation of Fisher transform sample partial correlation coefficients $\psi_{ij}$:
\begin{equation}\label{corSep}
\begin{split}
\psi_{ij} \doteq &\frac{1}{\sqrt{(1-w_{X_{ij}}^2)(1-w_{Y_{ij}}^2)}}[
w_{XY_{ii}}w_{XY_{jj}} +w_{XY_{ij}}w_{XY_{ji}} + \\
&w_{X_{ij}}w_{Y_{ij}}(w_{XY_{ii}}^2 +w_{XY_{jj}}^2 +w_{XY_{ij}}^2 +w_{XY_{ji}}^2)/2 -\\
&\{w_{X_{ij}}(w_{X_{ij}}w_{XY_{ij}} +w_{XY_{ji}}w_{Y_{ij}}) + w_{Y_{ij}}(w_{XY_{ji}}w_{XY_{ii}} + w_{XY_{jj}}w_{XY_{ij}})\} 
].
\end{split}
\end{equation}
We shall remark that this excludes the perfect dependence case where $w_{X_{ij}}=1$ and $w_{Y_{ij}}=1$ in (\ref{corSep}). For instance, we consider unit weights for the matrix diagonal.
Furthermore, if we assume a diagonal dependence structure in $\Omega_{XY}^J$, the expression can be simplified by
\begin{equation}\label{icorSep}
\begin{split}
\psi_{ij} \doteq &\frac{
w_{XY_{ii}}w_{XY_{jj}}+ w_{X_{ij}}w_{Y_{ij}}(w_{XY_{ii}}^2 +w_{XY_{jj}}^2)/2}{\sqrt{(1-w_{X_{ij}}^2)(1-w_{Y_{ij}}^2)}}.
\end{split}
\end{equation}
We propose two estimators for $[\psi_{ij}]$:
\begin{enumerate}
\item Regression-based estimator (Reg-based): 
\begin{equation}\label{icorSepEst}
\begin{split}
\hat{\psi}_{ij} = &\frac{
\hat{w}_{XY_{ii}}\hat{w}_{XY_{jj}}+ \hat{w}_{X_{ij}}\hat{w}_{Y_{ij}}(\hat{w}_{XY_{ii}}^2 +\hat{w}_{XY_{jj}}^2)/2}{\sqrt{(1-\hat{w}_{X_{ij}}^2)(1-\hat{w}_{Y_{ij}}^2)}}.
\end{split}
\end{equation}
where $\hat{w}_{X_{ij}}$ and $\hat{w}_{Y_{ij}}$ are estimates for $w_{X_{ij}}$ and $w_{Y_{ij}}$ respectively. These can be found using eq. \eqref{regInv} on the initial iteration of the ADMM Algorithm \ref{algJGL}. 

Moreover, $\hat{w}_{XY_{ii}}$ and $\hat{w}_{XY_{jj}}$ can be computed by considering a regression-type partial correlation coefficient estimation. For instance, the expression for variable $i$ is defined by
$\hat{w}_{XY_{ii}} = \text{cor}(X_{ki}- X_{k,-i}\hat{\beta}_{X_{i,-i}}, Y_{ki}- Y_{k,-i}\hat{\beta}_{Y_{i,-i}})$, with regression coefficients $\hat{\beta}_{m_{i,-i}} = -(\hat{\Omega}_m)_{i,-i}/(\hat{\Omega}_m)_{i,i}$ for $m=X,Y$.

\item Regression-based simplified estimator (Reg-based-sim): 
\begin{equation}\label{icorSepEst2}
\hat{\psi}_{ij}  = \hat{w}_{XY_{ii}}\hat{w}_{XY_{jj}}(1-\hat{w}_{X_{ij}}^2)^{-1/2}(1-\hat{w}_{Y_{ij}}^2)^{-1/2}. 
\end{equation}
for same regression-based estimators of $\hat{w}_{XY_{ii}}$ and  $\hat{w}_{XY_{jj}}$ as well as  partial correlation estimators  $\hat{w}_{X_{ij}}$ and $\hat{w}_{Y_{ij}}$  as defined in the Reg-based estimator, thus using the main terms in eq. \eqref{icorSep}.

\end{enumerate}
%


\section{Selection of tuning parameters}\label{tuning}
\subsection{Combination of two regularization parameters }\label{tuning1}
The joint estimation problem described in Section \ref{GM} requires the selection of two regularization parameters: $\lambda_1$ (sparsity) and $\lambda_2$ (similarity), and the combination of the two characterizes the estimated network sizes (both common network and differential network).
In terms of the differential network $\Omega_d = \Omega_Y - \Omega_X$, the same number of non-zero estimated differential elements, say $s_d = \sum_{i<j} I(\hat{\Omega}_{Y_{ij}} - \hat{\Omega}_{X_{ij}}\neq 0)$, can be achieved for many different combinations of the two parameters. For example, if we want to estimate $s_d=50$ differential edges, these can be found by the two extremes: (1) setting $\lambda_2=0$ and selecting $\lambda_1$ such that we have a maximum of  $50$ edges for each graphs ($\sum_{i<j} I(\hat{\Omega}_{X_{ij}}\neq 0 \,\, \text{or}\,\, \hat{\Omega}_{Y_{ij}}\neq 0) = 50$); (2) setting $\lambda_1=0$ and find $\lambda_2$ such that $s_d=50$ (in this case, $\sum_{i<j} I(\hat{\Omega}_{X_{ij}}\neq 0 \,\, \text{or}\,\, \hat{\Omega}_{Y_{ij}}\neq 0) = p(p-1)/2$).  In between, there are infinitely many combinations of $\lambda$'s that reach the same value for $s_d$ with the total number of edges being an upper bound for the number of differential edges. 

In \cite{Caballe2016} we discussed different ways of choosing sparsity penalization parameters that encourage certain
network characteristics, i.e. clustering structure or connectivity of the estimated networks. These could also be 
applied for the joint estimation algorithm once  the parameter $\lambda_2$ is fixed. Furthermore, here we want to propose an alternative procedure that transforms the problem of selecting regularization parameters  $\lambda_1$ and $\lambda_2$ to setting the desired expected proportion of false positive edges (EFPR) using parameters $\alpha_1$ (sparsity) and $\alpha_2$ (similarity).  This is possible to do directly (no resampling) and fast for the nature of the ADMM recursive algorithm presented in Section \ref{GM}, that,  for every iteration, obtains a dense estimation of the precision matrices before thresholding (see step 4 and 6). By having the whole dense matrix we can approximate a distribution that represents estimated coefficients whose true values are zero. 
In contrast, for graphical lasso algorithms in which the thresholding step is applied row by row using a regression based approach \citep{Friedman2007}, the EFPR is commonly controlled using subsampling methods  \citep{Meinshausen2010}, which increases considerably the computational cost.

\subsection{Selection of the expected false positive rate}\label{tuning2}
We define the sets $S_m = \{ (i,j),\, i<j: \,\Omega_{m_{ij}} =0\}_{m=X,Y}$ and the set $S_C = \{ (i,j),\, i<j:\, \Omega_{X_{ij}}-\Omega_{Y_{ij}} =0\}$ with $|S_m| = \text{Card}(S_m)$ and $|S_C| = \text{Card}(S_C)$. The objective is to set significance levels such that 
 \[
    \left\{
                \begin{array}{ll}
  \alpha_m = |S_m|^{-1}\sum_{(i,j) \in S_m} \mathbf{E}[I(\hat{\Omega}_{m_{ij}} \neq 0)], \, \, \, \, \, m= X, Y, \\
\alpha_2 =  | S_X\cap S_Y|^{-1}\sum_{(i,j) \in S_X\cap S_Y} \mathbf{E}\left[    \frac{I(\hat{\Omega}_{Y_{ij}} -\hat{\Omega}_{X_{ij}} \neq 0)}{ I(\hat{\Omega}_{Y_{ij}} \neq 0 \,\,\,\cup \,\,\,\hat{\Omega}_{X_{ij}} \neq 0)}\right].
                \end{array}
              \right.
  \]
Here we use $\alpha_X = \alpha_Y = \alpha_1$ by default.

The main characteristic of our proposed procedure is the adjustment of the penalization parameters $\lambda_1$ and $\lambda_2$ in every iteration $t \in 1:T$ of the joint estimation algorithm depending on the significance levels $\alpha_1$ and $\alpha_2$ as well as the estimated precision matrices $\hat{A}_Y^{(t)}$ and $\hat{A}_X^{(t)}$ (which are described in step 6 of Algorithm \ref{algJGL}). 
We define $\hat{A}_D^{(t)} = [v_{ij}(\hat{A}_{Y_{ij}}^{(t)}-\hat{A}_{X_{ij}}^{(t)})^{1/2} ]_{i<j}$ as a vector of size $p(p-1)/2$ with the standardized estimated partial correlation coefficient differences at iteration $t$ and $\hat{A}_J^{(t)} = \{ [\hat{A}_{X_{ij}}^{(t)}]_{i<j}, [\hat{A}_{Y_{ij}}^{(t)}]_{i<j} \}$ as a vector of size $p(p-1)$ with the updated estimated partial correlation coefficients in both populations. We assume that coefficients $\hat{A}_{J_{h}}^{(t)}$ with $h \in S_m$ follow a normal distribution with mean zero and variance $\sigma_{1(t)}^2$ and that 
coefficients $\hat{A}_{D_{h}}^{(t)}$ with $h \in S_C$ follow a normal distribution with also mean zero and variance $\sigma_{2(t)}^2$. In both cases, normality assumption is justified for sufficiently large sample size in our simulated data study and the analysis is presented in the supplementary material.

The pairs of variable where similarity or sparsity conditions hold are unknown without any prior information and variance parameters cannot be estimated by their sample estimators using elements in such defined sets. However, we assume that (i) most of the coefficients are zero in any of the two matrices (strong sparsity) and (ii) most of the coefficients are equal between the two precision matrices (strong similarity). Hence, we  propose to use robust estimators for the variances $\sigma_{1(t)}^2$ and $\sigma_{2(t)}^2$  using all the partial correlation coefficients (differences). In this way, we reduce the importance of large coefficients which can be generated by true non-zero partial correlation coefficients.  Next, we describe three of the most popular estimators for the variance in the robust statistics literature which are fully described in \cite{Rousseeuw1993}:


\begin{enumerate}
\item Median absolute deviation around the median:
$$
\hat{\sigma}_{2(t)} = 1.4826\text{mad}(\hat{A}_D^{(t)})  \mbox{\hspace{1cm}}
\hat{\sigma}_{1(t)} = 1.4826\text{mad}(\hat{A}_J^{(t)}),
$$
where $\text{mad}(x)  = \text{median}( |x_i- \text{median}(x)|)$. 
\item Interquartile range:
$$
\hat{\sigma}_{2(t)} = \frac{\text{IQR}(\hat{A}_D^{(t)})}{1.349}, \mbox{\hspace{1cm}}
\hat{\sigma}_{1(t)} = \frac{\text{IQR}(\hat{A}_J^{(t)})}{1.349},
$$
where
$\text{IQR}(x)  = q(x)_{0.75} - q(x)_{0.25}$ with $\alpha$-quanitle $q(x)_\alpha$. 
\item Rousseeuw and Croux (RC) mad alternative:
$$
\hat{\sigma}_{2(t)} = 1.1926 \text{RCmad}(\hat{A}_D^{(t)})\mbox{\hspace{1cm}}
\hat{\sigma}_{1(t)} = 1.1926 \text{RCmad}(\hat{A}_J^{(t)}),
$$
where $\text{RCmad}(x)  = \text{median}_i\{\text{median}_j|x_i-x_j|\}$. 
\end{enumerate}
For each iteration $t \in T$, another way to select the pair $[\lambda_1, \lambda_2]$ is by fixing significance levels $[\alpha_1, \alpha_2]$ and considering $[\lambda_1=Z_{\alpha_1}  \hat{\sigma}_{1(t)},$ $\lambda_2=Z_{\alpha_2'}  \hat{\sigma}_{2(t)}]$ where $Z_{\alpha}$ is the upper $\alpha$ critical value for the standard normal distribution and
$\alpha_2'$ is determined by 
$$
\alpha_2 \doteq (p_1p_3)/(p_1p_3 + p_2(1-p_3)),
$$ 
with
$p_1 = \Pr( |z_1| > Z_{\alpha_2'}/2 + Z_{\alpha_1}/\sqrt{2} \,\, \cup \,\, |z_2| > Z_{\alpha_2'}/2 + Z_{\alpha_1}/\sqrt{2} | \,\,|z_1-z_2|> Z_{\alpha_2'} )$,
$p_2 = \Pr( |z_1 + z_2|/2 >  Z_{\alpha_1}/\sqrt{2}| \,\,|z_1-z_2|< Z_{\alpha_2'} )$ and $p_3 = \Pr( |z_1-z_2|> Z_{\alpha_2'} )$ using independent standard normal distributed random variables $z_1$ and $z_2$. We approximate $\alpha_2'$ by Monte Carlo. 
%
Default values for  $\alpha_1$ and $\alpha_2$  as $0.01$ or $0.05$ could be used. Note that the estimated variances are found using $\approx p(p-1)/2$ coefficients, but in case $p$ is very small, Student's t-quantiles could be used instead of $Z_{\alpha}$  to approximate the CI.  

\subsection{Uncertainty in the differential network}\label{uncertainty}
Differential network estimators incorporate the variability of the two individual estimated networks and tend to be much more uncertain that the underlying estimated common network. 
Here we propose to perform a permuted samples based approach to assess the uncertainty in the number of estimated differential edges. We permute the data as follows to ensure that the dependence structure between datasets is maintained:
 $[(Z_1^{\pi_1},\ldots, Z_n^{\pi_n}),(Z_1^{\bar{\pi}_1},\ldots, Z_n^{\bar{\pi}_n})]$ where $\bar{\pi}_i=1-\pi_i$ and
\begin{equation}
\begin{cases}
Z_i^{\pi_i} = X_i \mbox{ if } \pi_i = 0,\\
Z_i^{\pi_i} = Y_i \mbox{ if } \pi_i = 1,
\end{cases}
\end{equation}
with $\Pr(\pi_i = 1) = 0.5$.  Given the new permuted data, a weighted fused graphical lasso estimate can be found by solving eq. \eqref{DAN} using the same combination for $\lambda$'s as for the original estimate. By repeating this permutation and estimation process $T$ times, we can compute a confidence region for the number of estimated differential edges (distinguishing between the two populations) under the hypothesis of equality in the two precision matrices: $H_0: \Omega_X = \Omega_Y$.

\section{Overestimation of triangular motifs}\label{triangle1}
\subsection{Problem and toy example}
We discovered that the overestimation of triangles is a major issue here. If there are 3 nodes A,B,C and we already know that pairs A,B and  A,C, are connected, then a connection between B and C is more often falsely predicted than expected. The reason for this is that the ADMM-type algorithm presented in Algorithm \ref{algJGL} uses a regularization for the eigenvalues $[D_{jj}]$ of the covariance/correlation matrix to approximate its inverse denoted by $[\tilde{D}_{jj}]$ (see eq. \eqref{regInv}). It can be proved that when $D_{jj}\gg \sqrt{\rho/n}$ then $\tilde{D}_{jj} \approx 1/D_{jj}$ and when $D_{jj}\leq c \sqrt{\rho/n}$ then $\tilde{D}_{jj}\approx \tilde{c} \sqrt{n/\rho}$. In the second such scenario, which happens when $n$ is small in comparison to $p$,  the estimated coefficients are biased.

\begin{figure}[ht]
\begin{center}
 \begin{tabular}{cc}
    \subfloat[n=25, weight=25]{\includegraphics[scale=0.22]{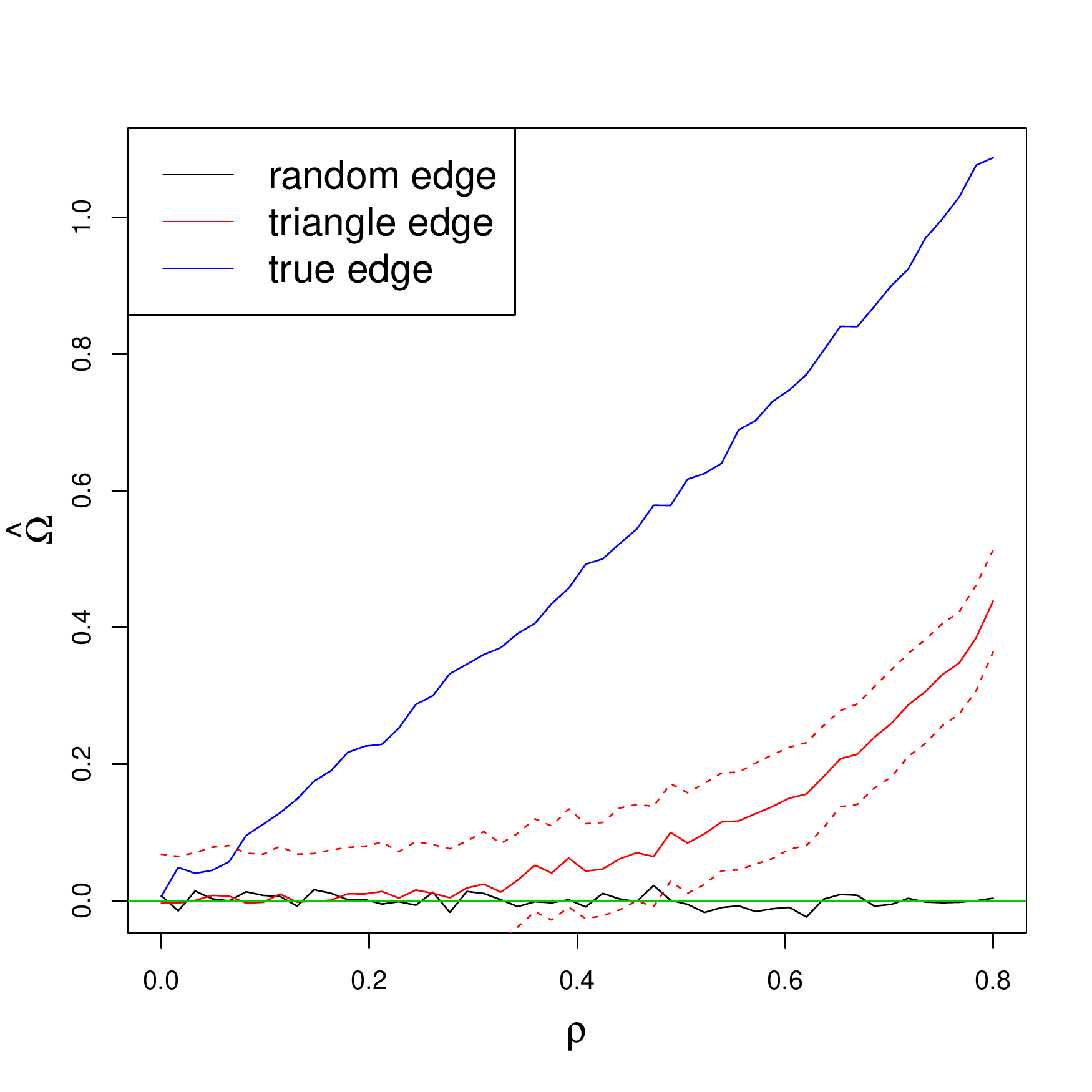}}&
    \subfloat[n=100, weight=100]{\includegraphics[scale=0.22]{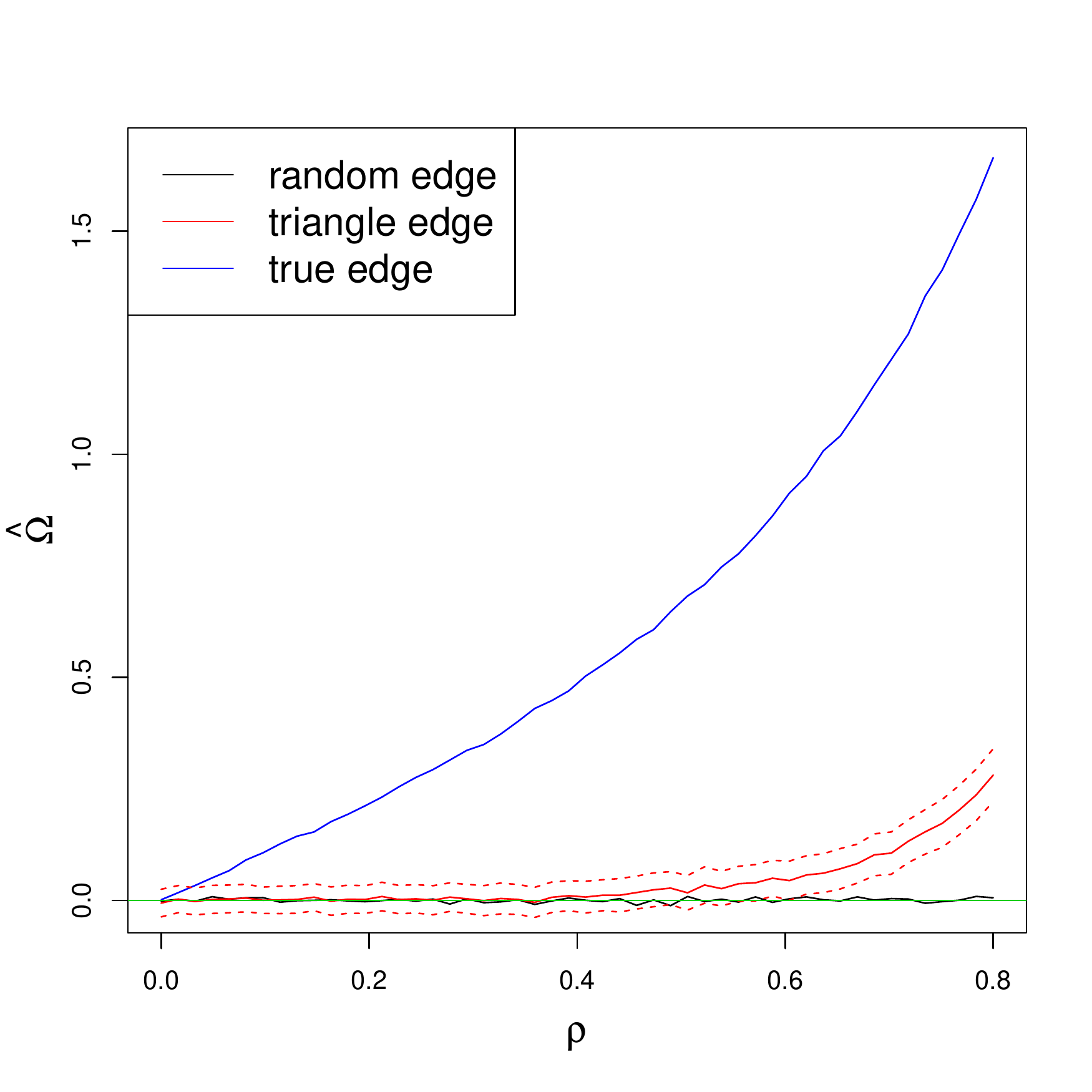}}\\
    \subfloat[n=500, weight=500]{\includegraphics[scale=0.22]{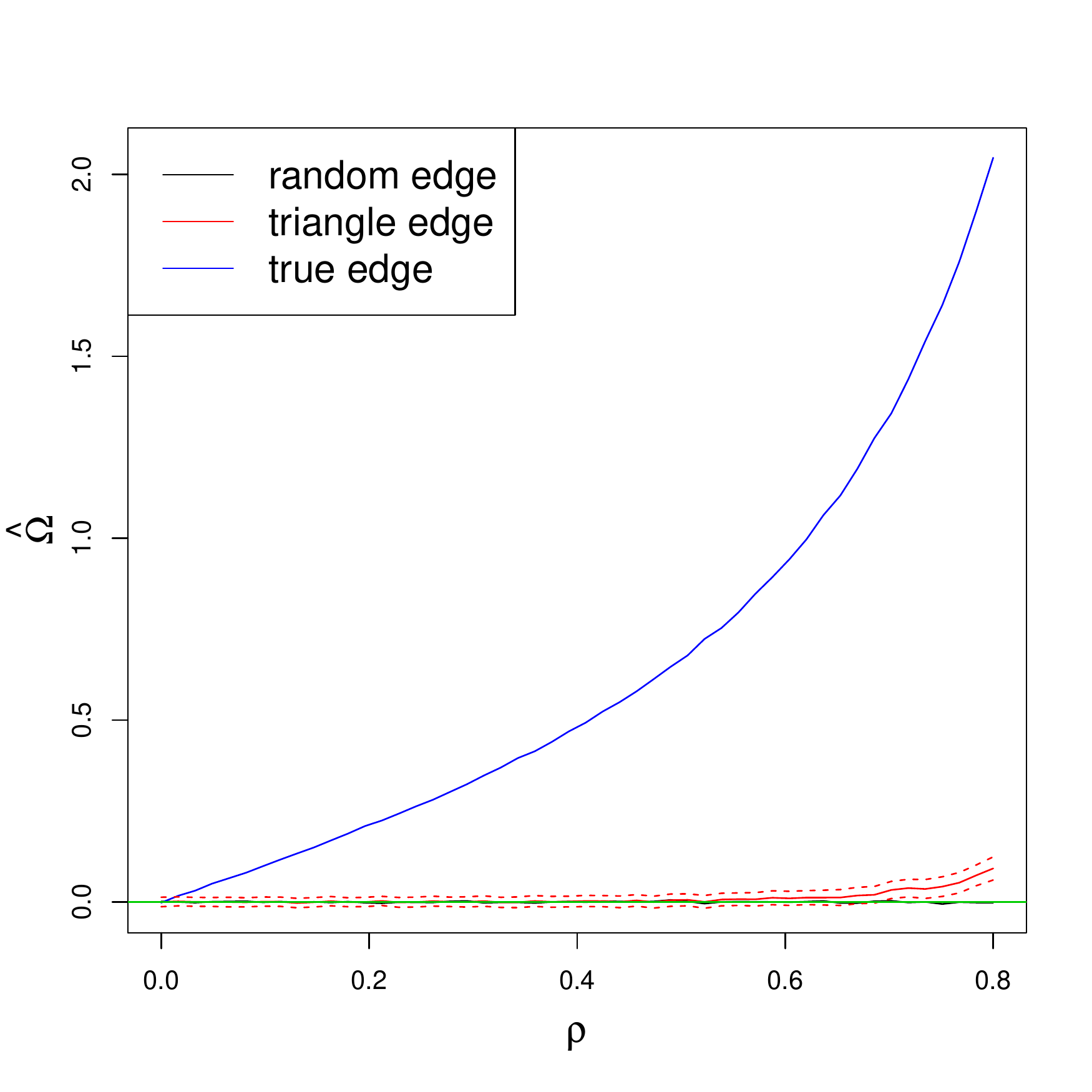}}&
    \subfloat[n=500, weight=1]{\includegraphics[scale=0.22]{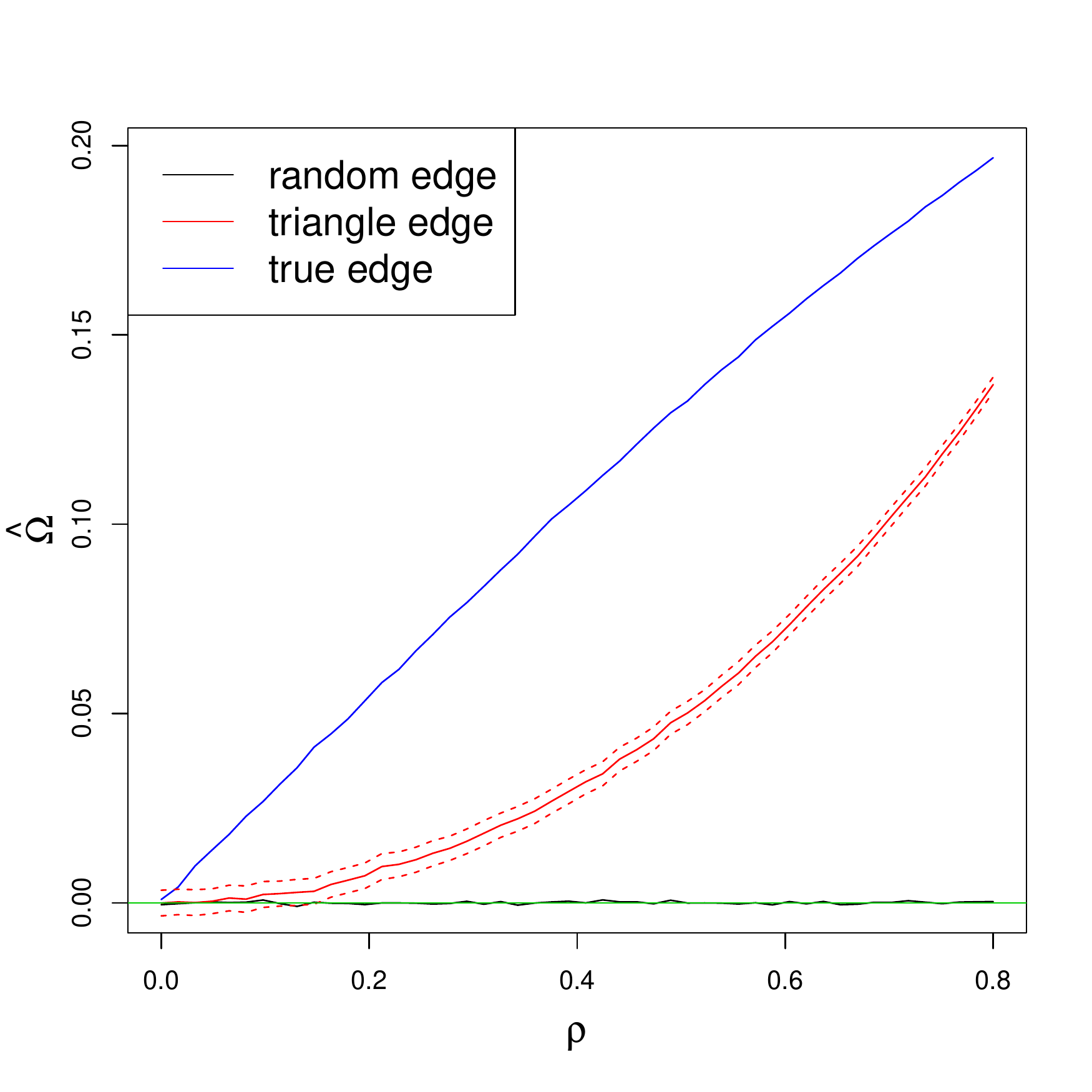}}\\
 \end{tabular}
 \caption{{\footnotesize average precision matrix estimated value for true edge (blue), false triangle edge (red) and false edge (green).}}
\label{fSim2C6}
\end{center}
\end{figure}

We illustrate this using a toy graph structure example described by:
$$
G_x: 1\longleftrightarrow 2, 1\longleftrightarrow 3, 4\longleftrightarrow \emptyset;
$$
hence, here the edge $2\longleftrightarrow 3$ is the one missing to do a triangle. Assuming that the edges $1\longleftrightarrow 2$ and $1\longleftrightarrow 3$ have the same strength, we can express the correlation matrix and its inverse by
$$
R_{X} = \begin{pmatrix} 1 & & & \\ \rho & 1 & & \\ \rho& \rho^2 &1 &\\0 &0 & 0 & 1 \end{pmatrix}, \mbox{\hspace{1cm}} R_X^{-1} = \Omega_{X} = \begin{pmatrix} 1 & & & \\
-\frac{\rho-\rho^3}{1-(2\rho^2-\rho^4)} & 1& & \\
-\frac{\rho-\rho^3}{1-(2\rho^2-\rho^4)} &0 &1 &\\
0 &0 & 0 & 1 \end{pmatrix}.
$$
To show the behavior of the regularized precision matrix estimator defined at \eqref{regInv} we
simulate data from a multivariate normal distribution with mean vector equal to zero and covariance matrix equal to $R_X$. In Figure \ref{fSim2C6} we show the trend of $-\hat{\Omega}_{12}$ (true edge), $-\hat{\Omega}_{14}$ (false edge) and $-\hat{\Omega}_{23}$ (false triangle edge) for different sample sizes and over 1000 simulations. We shall see that the $\hat{\Omega}_{12}$ is shrunk towards zero for small $n$ as expected, also  $\hat{\Omega}_{14}$ is centered at zero as expected but $\hat{\Omega}_{23}$ is biased. The true $\Omega_{23}=0$, but for $\rho$ large enough the expected value of estimated $\hat{\Omega}_{23}$ is different from zero. 

\cite{Danaher2014} make an additional consideration to the formula in (\ref{regInv}). They suggest to use $n$ as a vector with the weights of the classes and default values equal to $1$ for all classes. The reason is that even though using $n$ as sample size reduces the bias, it also gives much larger variances for edges equal to zero than using the default weights and, therefore, it can produce more false positive edges. However, using weights equal to $1$ has as main problem precisely the detection of false positive triangular motifs. This is reflected in  Figure \ref{fSim2C6}(d), where the regularized inverse produces even a larger bias for the false triangle edge than in previous cases.   


\subsection{Reducing overestimation of triangular motifs}
Here we focus on the hypothesis testing problem defined by H$_0$: not a triangle and H$_1$: triangle.  The null hypothesis  holds if any of the three partial correlation coefficients associated to the three edges that make the triangle is zero. Note that this contains multiple scenarios under $H_0$ and finding a reliable null distribution is not suitable without prior information.  For instance, there are 7 configurations of the graph structure which can be considered under the null hypothesis (1 for all zero values, 3 for only one non-zero value and also 3 for two non-zero values) and in each one of them the correlation between estimated coefficients is different. 
Since the ADMM algorithm does find well conditionally dependence structures, we simplify the testing problem  by assuming that under $H_0$ two edges are present, thus that there is a pair of variables, say $(i,j)^*$, that is conditionally independent to the rest (the missing one to complete the triangle). 

To test the existence of this motifs structure we employ the scaled inverse of the $3\times 3$ correlation matrix (which determines the partial correlation matrix) that involve the three nodes in the triangle. We use the Fisher transformation function $g: \mathbb{R}  \to \mathbb{R}$, $g(z) = \log\{(1+z)/(1-z)\}/2$ on the estimated partial correlation coefficient $\hat{w}_{ij}$ such that $g(\hat{w}_{ij^*})$ is approximately normal with mean value zero and variance $n-5$ \citep{Fisher1924}. In case the pair $(i,j)^*$ was known, then the p-value of the test would be calculated by
\begin{equation}\label{probNorm}
\text{p-val} = P(|Z|  \geq |g(\hat{w}_{ij})|) \doteq 2 - 2\Phi\left( \sqrt{n-5} (|g(\hat{w}_{ij})|)\right),
\end{equation}
where $Z$ defines the standard normal with cumulative distribution $\Phi$.
We approximate a p-value for the test in case the position of the pair $(i,j)^*$ is  unknown by applying \eqref{probNorm} on the minimum estimated coefficient in absolute value $g' = \text{min}\{ |g(\hat{w}_{(ij)})|\}_{i<j}$. This results to a conservative p-value: for example if pair $(1,2) = (i,j)^*$, then  it is immediate to see that 
$$
\Pr(|Z|  \geq |g(\hat{w}_{12})| \cup|Z|  \geq |g(\hat{w}_{13})| \cup|Z|  \geq |g(\hat{w}_{23})|)\geq \Pr(|Z|  \geq |g(\hat{w}_{12})|).
$$
For large sample sizes (or very large true non-zero partial correlation coefficients), then the equality holds. 

Here, we assess the weakest edges of all the observed triangular structures separately and we eliminate those with small p-values (default threshold equal to $\alpha_1$ as described in Section \ref{tuning}). In case one edge is tested more than once, we only count its smallest p-value. Nevertheless, multiple testing correction and another interpretation for overlapping triangles could be used instead.

The hypothesis testing problem studied in this section could also be applied to data in the field of decomposable graphs \citep{Giudici1999, Green2013} in which networks are totally described by the presence of triangular motifs.

\section{Simulated data analysis}\label{SEC4}
\subsection{Models used to generate the data}\label{powerLaw}
We generate data from multivariate normal distributions with zero mean vector and several almost-block diagonal precision matrices, where each block (or cluster) has a power-law underlying graph structure (defined below) and there are some extra random connections between blocks. The non-zero partial correlation coefficients are simulated by
\begin{equation}
\Omega^{(0)}= [\omega_{ij}^{(0)}], \mbox{\hspace{0.5cm}} \omega_{ij}^{(0)} = \left\{ \begin{array}{r l l}
      & \text{Unif}(0.5,0.9) & \mbox{if $E_{ij} = 1$ with prob$=0.5$ } ;\\
      &  \text{Unif}(-0.5,-0.9) & \mbox{if $E_{ij} = 1$ with prob$=0.5$ }; \\
      &   0 & \mbox{if $E_{ij} = 0$}.
         \end{array} \right.
\label{eqSM11}
\end{equation}
Then, we regularize $\Omega^{(0)}$ by $\Omega^{(1)} = \Omega^{(0)} +  \delta I$, with $\delta$ such that the condition number of  $\Omega^{(1)}$ is less than the number of nodes, so obtaining a positive definite matrix \citep{Ai2011}. 
Initially we consider  $\Omega_X^{J}=\Omega_Y^{J}=\Omega^{(1)}$. However, we also include differential edges using additional block diagonal structures. For instance, we use two block diagonal structures $D_X$, $D_Y$ so that we merge $[D_X,I]$ to $\Omega_X^{J}$ and $[I,D_Y]$ to  $\Omega_Y^{J}$. 

As for \cite{Wit2015} we define  $\Omega_{XY}^{J}$ by a diagonal matrix. Nevertheless, we consider that these elements in the diagonal can be different, for instance we use $\Omega_{XY_{ii} }^{J} = 0.6$ for $\lfloor p/2\rfloor$ diagonal elements (chosen randomly) and $\Omega_{XY_{ii} }^{J}= 0$ for the other $\lceil p/2\rceil$. 

Power-law networks assume that the variable $p_k$, which denotes the fraction of nodes in the network that has degree $k$, follows a power-law distribution
$$
p_k =  k^{-\alpha}\varsigma(\alpha)^{-1},
$$
where $k \geq 1$, a constant $\alpha>0$ and the normalizing function $\varsigma(\alpha)$ is the Riemann zeta function. Following \cite{Peng2009}, $\alpha = 2.3$ provides a good representation of biological networks. 

\begin{figure}[h]
\begin{center}
 \begin{tabular}{ccc}
\subfloat[Network example  p=200]{\includegraphics[scale=0.27]{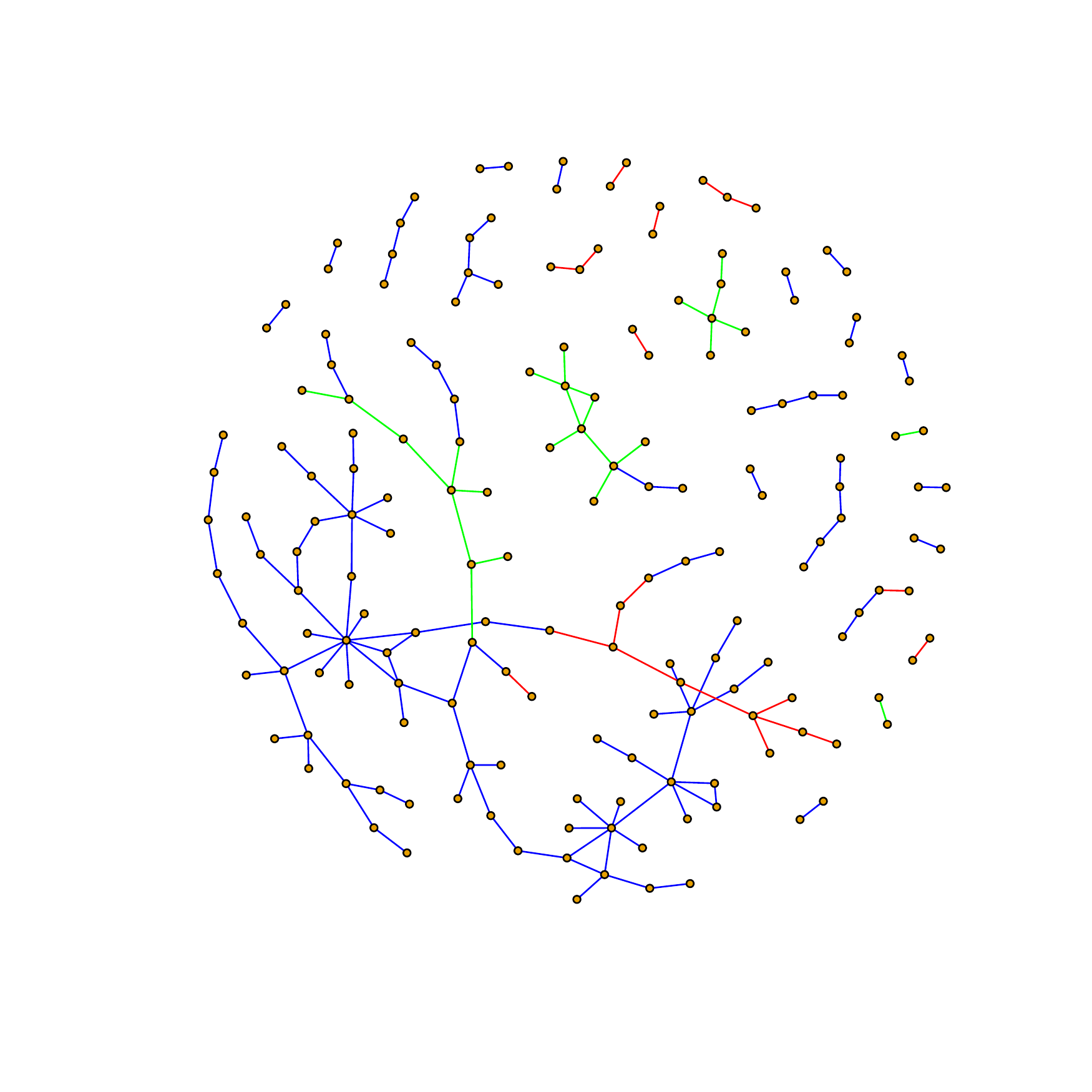}}&
\subfloat[Network example  p=300]{\includegraphics[scale=0.27]{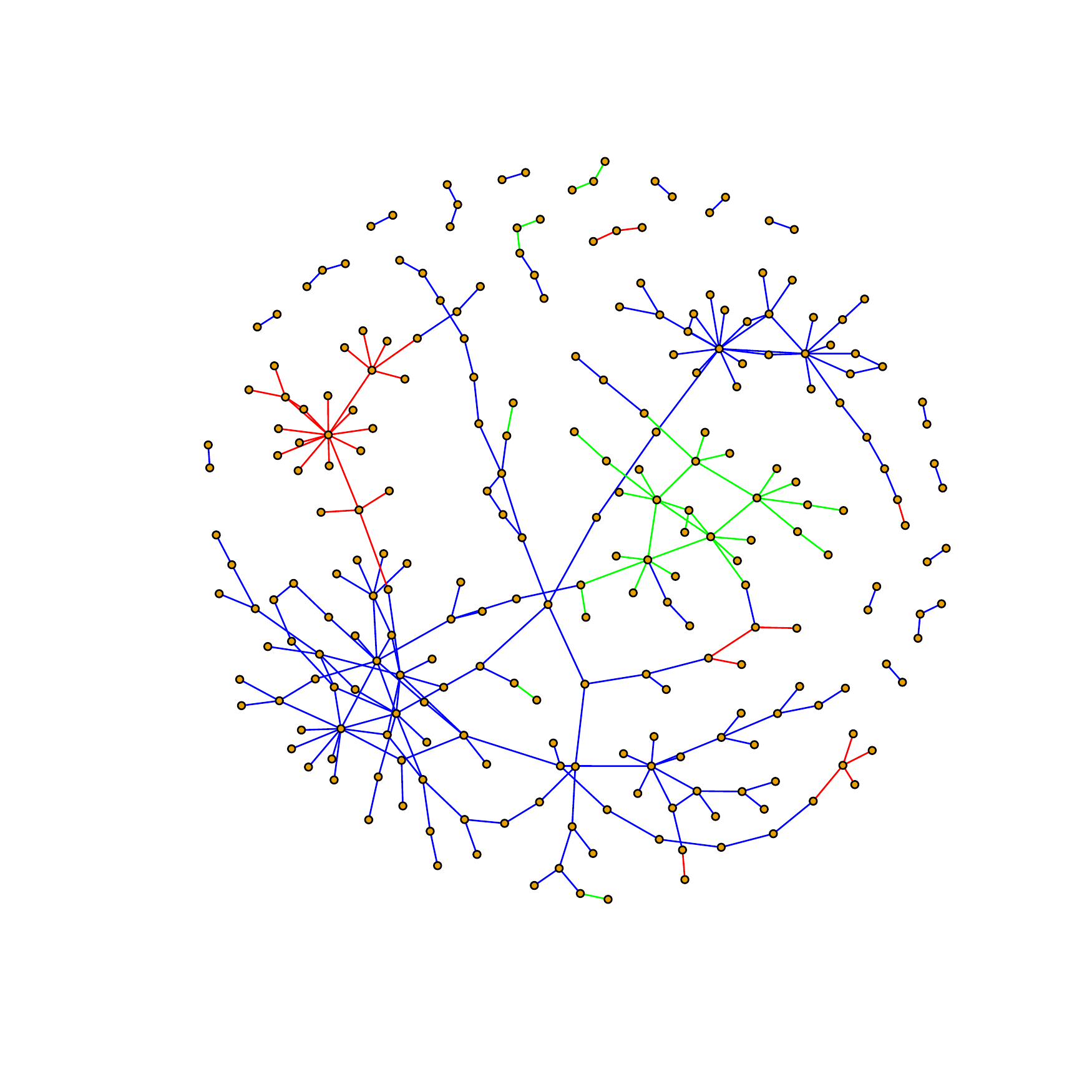}}&
\subfloat[Network example  p=400]{\includegraphics[scale=0.27]{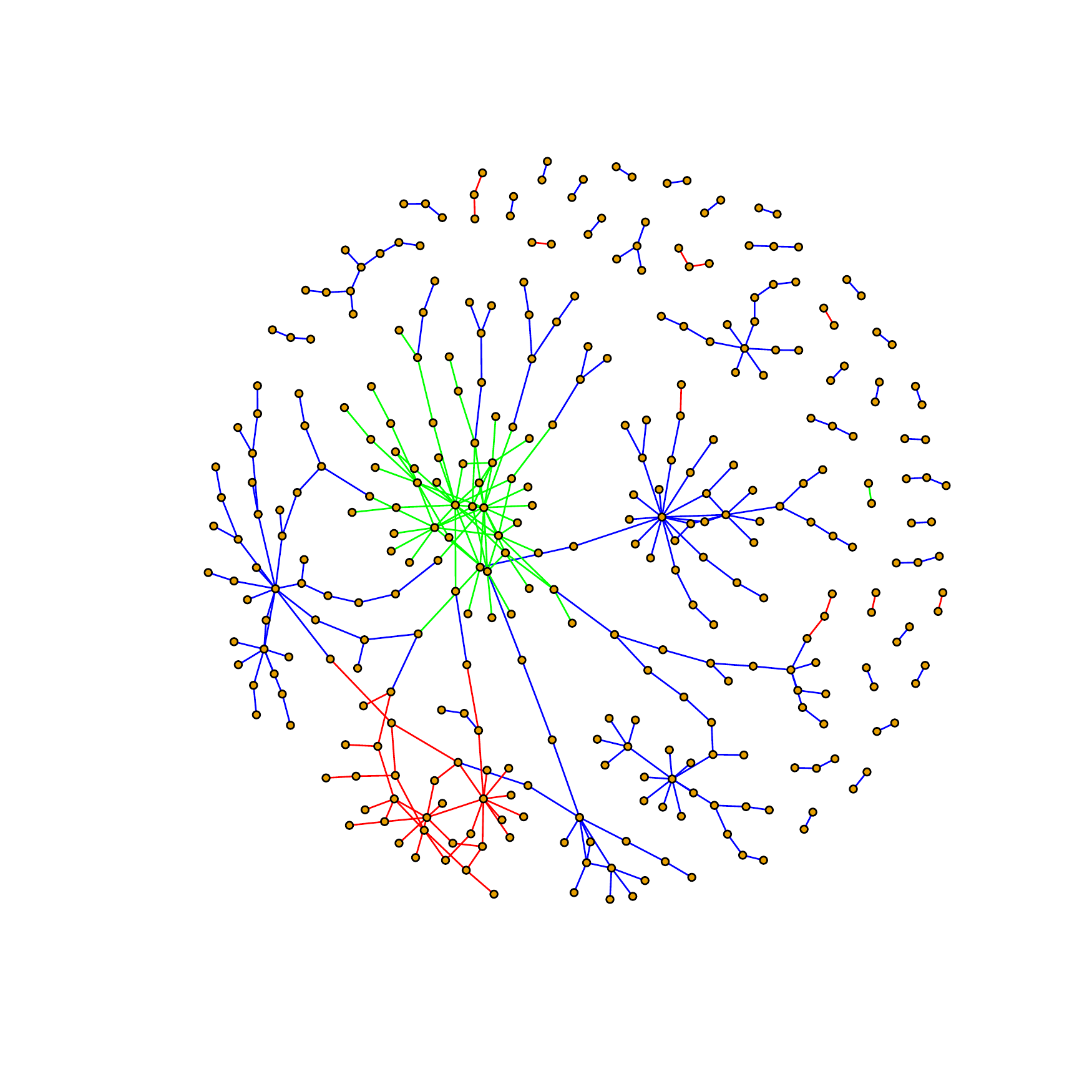}}
\end{tabular}
\caption{\footnotesize{Graph structure examples. Green edges are zero edges in the second population and non zero in the first population. Red edges are zero in the first population and non-zero in the second population. Finally, blue edges are non-zero and equal in both conditions.}}\label{fSim2C3}
\end{center}
\end{figure}

We generate datasets with several dimension sizes $p \approx $ 200, 300, 400 and sample sizes $n = $ 25, 100, 250, 500. In Figure \ref{fSim2C3} we present the graphical representation of some of the the generated networks.

\subsection{Differential network recovery}\label{DiffRecSim}
In this section we  focus on the recovery of differential edges by using two joint graphical lasso algorithms in the simulated datasets: FGL \citep{Danaher2014} and WFGL (proposed). In order to make the methods comparable we select estimated graphs (or $\lambda_1$ and $\lambda_2$) that have the same number of common edges and differential edges in the two approaches. We first select the pair $[\lambda_1, \lambda_2]$  for the WFGL approach by setting the expected false positive rate by the parameters  $[\alpha_1 = 0.05, \alpha_2 = 0.05]$ following the strategy proposed in Section  \ref{tuning}. Then we find $\lambda$'s such that the FGL graphs have the same sizes as WFGL. In total we use 200 iterations for each model, 4 different sample sizes $n=25, 100, 250, 500$ and three dimension sizes $p \approx $ 200, 300, 400. 

To compare the performance of the methods, we propose to use the Youden's index defined by  
$$
\text{YI}^M_\lambda = \text{TP}^{M}_\lambda - \text{FP}^{M}_\lambda, \mbox{\hspace{0.5cm}} \{M= \text{FGL}, \text{WFGL}\},
$$
where $\text{TP}^M_\lambda= \sum_{i<j} I[\hat{\Omega}_{X_{ij}}^M -\hat{\Omega}_{Y_{ij}}^M \neq 0, \Omega_{X_{ij}} - \Omega_{Y_{ij}} \neq 0)]$ and $\text{FP}^M_\lambda = \sum_{i<j} I[\hat{\Omega}_{X_{ij}}^M -\hat{\Omega}_{Y_{ij}}^M \neq 0, \Omega_{X_{ij}} - \Omega_{Y_{ij}} = 0)]$ are the number of true positives and false positive of the estimated differential graphs with $\lambda=[\lambda_1,\lambda_2]$ and method $M$. Then we compute
$$
\delta = \text{YI}^{\text{WFGL}}_\lambda - \text{YI}^{\text{FGL}}_\lambda,
$$
which defines the Youden's index differences between the two methods to estimate the joint networks. In Table \ref{tabPOWER2} we present the average difference (with a t-test p-value) and also the average sign of the differences $\delta$ (with a Wilcoxon test p-value). The proposed method that assumes a dependence structure achieves better TP-FP ratios for the differential network than the original FGL in most of the models when $n$ is fairly large. For small $n$, there are no significant differences between the two algorithms even when there exists a dependence structure in the data. 

\begin{table}[h]
\small
\begin{center}
\caption{ {\footnotesize Youden Index differences between WFGL and FGL algorithm.} } 
\begin{tabular}{ l r r   r r  r r rr rr}
&\multicolumn{2}{c}{p= 200} &&\multicolumn{2}{c}{p=300}&& \multicolumn{2}{c}{p=400}\\
\cline{2-3}\cline{5-6}\cline{8-9}
$n$  & $\bar{\delta}$ (p-val) & $\bar{sgn(\delta)}$ (p-val) &&$\bar{\delta}$ (p-val) & $\bar{sgn(\delta)}$ (p-val) &&$\bar{\delta}$ (p-val) & $\bar{sgn(\delta)}$ (p-val)\\
\hline
25   &.18 (0.07) &.06 (0.09) && .04 (0.48) & .01 (0.50) &&  .12 (0.11) & .05 (0.11) \\
100 &.25 (0.04) &.07 (0.06) && .16 (0.10) & .04 (0.20) &&  .26 (0.03) & .08 (0.06) \\
250 &.26 (0.02) &.10 (0.02) && .27 (0.05) & .09 (0.04) &&  .32 (0.02) & .11 (0.01) \\
500 &.24 (0.05) &.08 (0.05) && .15 (0.12) & .06 (0.07) &&  .19 (0.07) & .07 (0.08) \\
\end{tabular}
\label{tabPOWER2}
\end{center}
\end{table}

Even though the correction for dependent datasets does not improve in great measure the differential network recovery rates,  assuming that differential edges can occur with same probability independently of the values $[\psi_{ij}]$ produces a fairer procedure in which edges with high correlation have similar chances to be recovered as edges with low correlation. 

\begin{figure}[ht]
\begin{center}
\includegraphics[width=12cm,height=8cm]{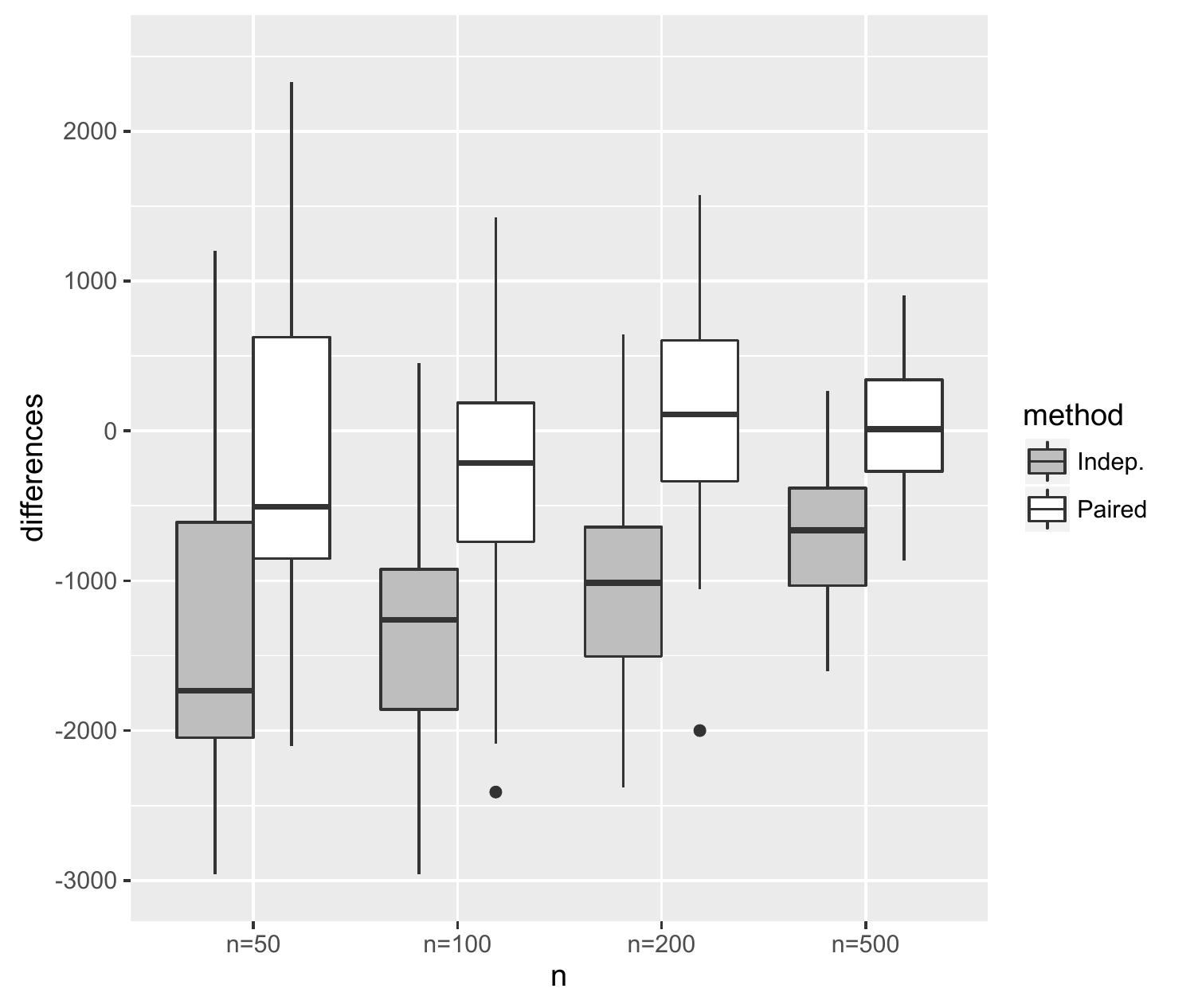} \caption{\footnotesize{Differences between $\Omega_d$ average ranks among large $\psi_{ij}$ and small  $\psi_{ij}$ over 50 simulations in the first iteration of the ADMM algorithm by correcting/non-correcting for dependence structure. }}\label{fdiffOmega}
\end{center}
\end{figure}

We show this using the model defined by a dimension $p=300$ and several sample sizes. We separate pairs of variables $(i,j)$ in two groups: $L= \{(i,j): \psi_{ij}<0.1\}$ and $U=  \{(i,j): \psi_{ij}>0.1\}$. For all pairs $(i,j)$, we compute  $h_{ij} = v_{ij}[|(\hat{\Omega}_Y)_{ij}-(\hat{\Omega}_X)_{ij}|]$ using $v_{ij}=1$ (Indep.) as well as $v_{ij}=(2-2\hat{\psi}_{ij})^{-1/2}$ (paired) with $[\hat{\psi}_{ij}]$ estimated by the Reg-based-sim method discussed in Section \ref{pairedWeights} (see Section 1 in the supplementary material for comparison between $\psi$ estimators). 
Then we rank the values $h_{ij}$ and we denote them by $k_{ij}$ such that $k_{sl}=1$ for $sl = \arg\max\limits_{(i,j)} h_{ij}$  and $k_{sl}=p(p-1)/2$ for $sl = \arg\min\limits_{(i,j)} h_{ij}$. In Figure \ref{fdiffOmega} we show the differences of the average ranks in the two groups, i.e. $|L|^{-1}\sum_{(i,j) \in L} k_{ij} - |U|^{-1}\sum_{(i,j) \in U} k_{ij}$. We can see that the independent method encourages recovery of differential edges with small $\psi_{ij}$ (seen in the plot  by large negative rank differences) and this bias is corrected  by the dependent data adjustment, which for relatively large sample size gives very similar ranks in the two groups. 

\subsection{Evaluation of tuning parameter selection}\label{NetworkRecSim}
\begin{figure}[ht]
\begin{center}
 \begin{tabular}{cc}
    \subfloat[n=25]{\includegraphics[scale=0.31]{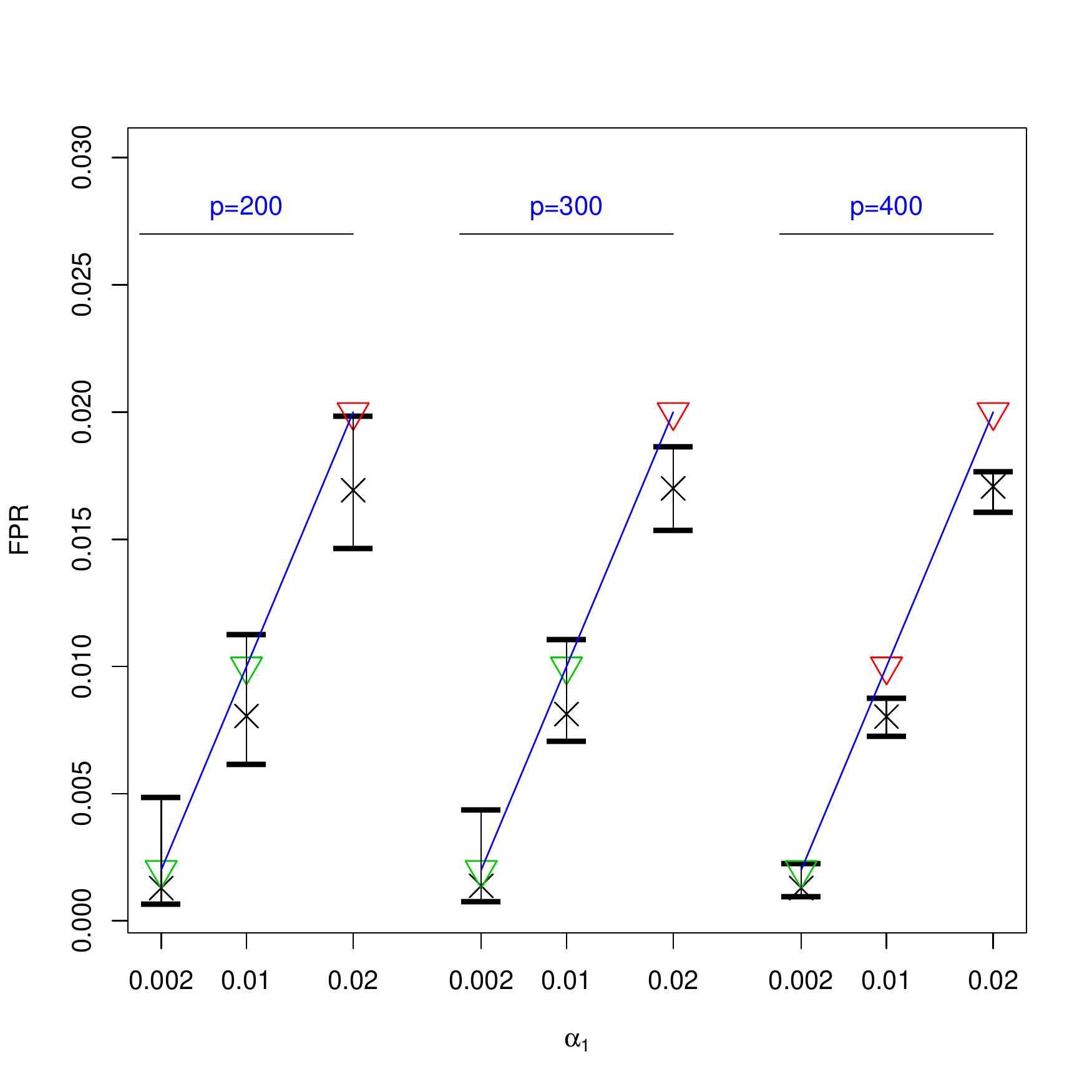}}&
    \subfloat[n=100]{\includegraphics[scale=0.31]{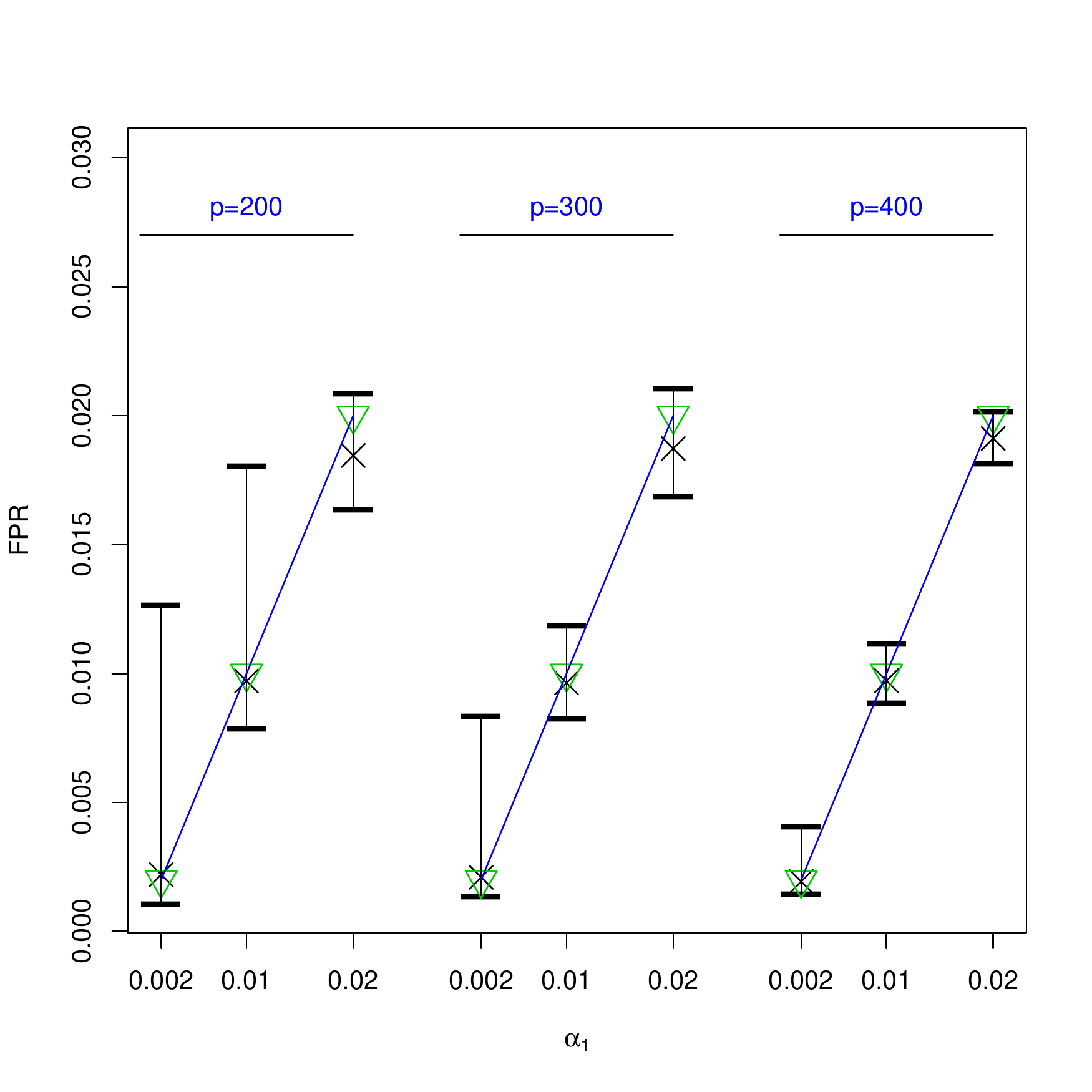}}\\
    \subfloat[n=250]{\includegraphics[scale=0.31]{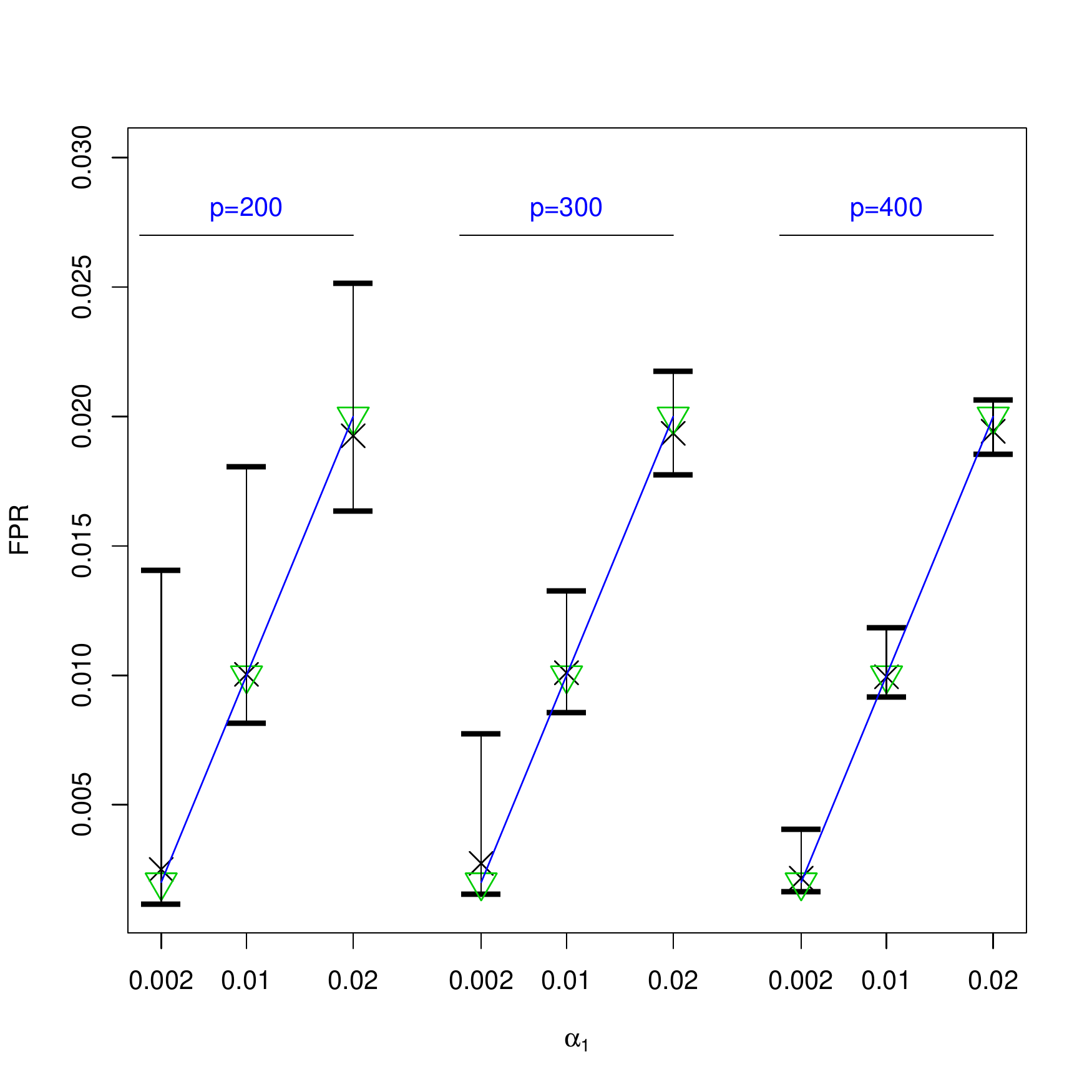}}&
    \subfloat[n=500]{\includegraphics[scale=0.31]{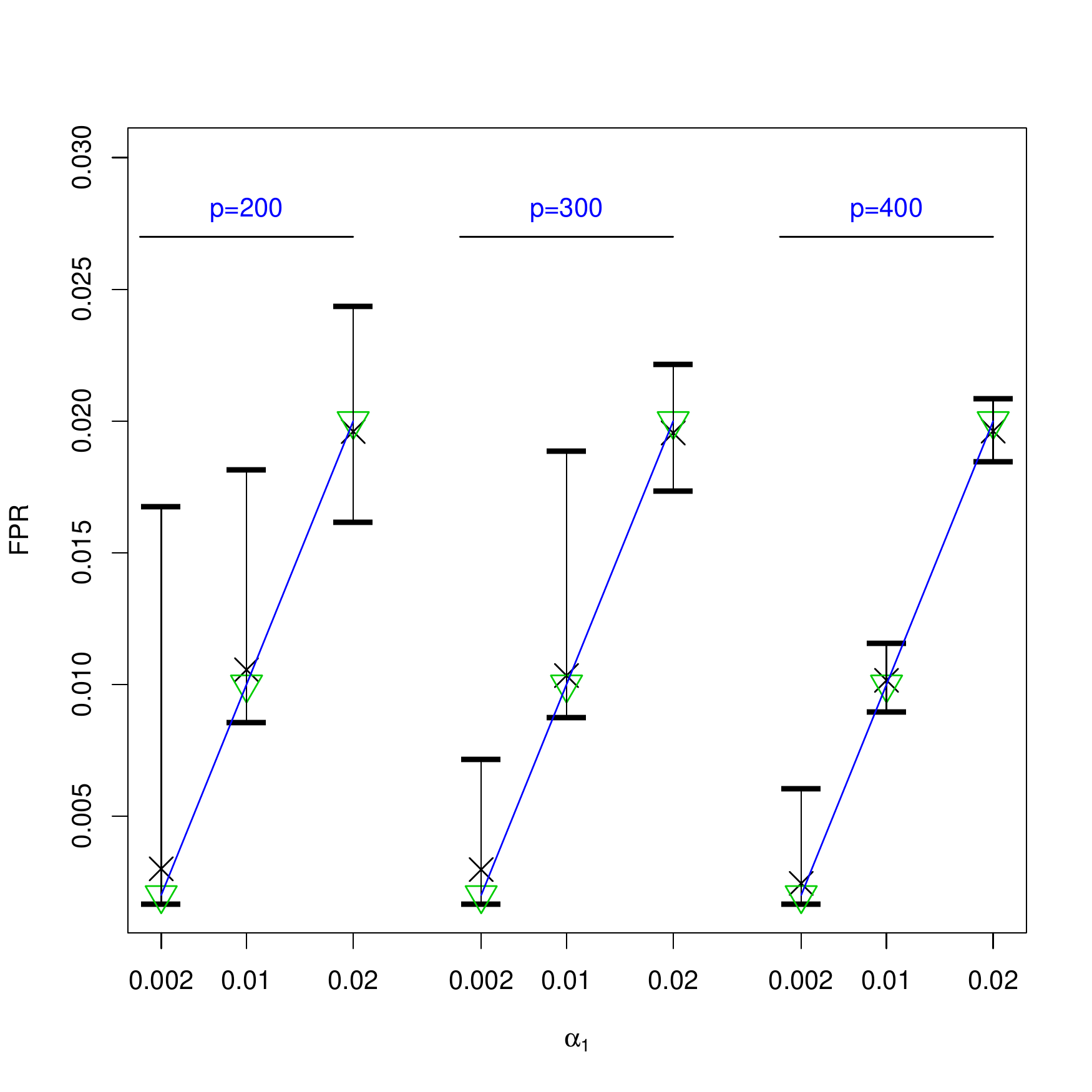}}
 \end{tabular}
 \caption{{\footnotesize FPR vs $\alpha_1$: average (cross) + CI is plotted together with the expected values (triangle). For visualization reasons, x-axis and y-axis are not in the same scale (i.e. $2x : y$)}.}
\label{fSim2C7}
\end{center}
\end{figure}

In Figure \ref{fSim2C7} we compare the expected proportion of false positive edges determined by the value of $\alpha_1$ against the observed false positive rate  (with median and  $95\%$ confidence) using the RCmad estimator to approximate $\sigma_1$ (see Section 2 in the supplementary material for comparison between $\sigma$ estimators). To draw the confidence interval we replicate the procedure in 100 simulated datasets for different sample sizes and dimension sizes. The approximated false positive rate is close to the true one, given by $\alpha_1$, and it is only for very small $n$ that the true value is not always included in the confidence interval. 
A similar analysis is applied to the other tuning parameter $\alpha_2$ in the supplementary material.

\subsection{Testing and removing triangle motifs}\label{triangle2}
As we discussed in section \ref{triangle1}, using the eigenvalue decomposition regularization forces a bias to some non existing edges in the true network. These ones are the missing edges to form closed structures. For example, triangle structures are the ones that suffer the most this bias. This is illustrated in Table \ref{tabPOWER16}  where we present the average TP-FP behavior for the weakest edge of estimated triangles for models with different sample sizes, dimensions sizes and significance levels $\alpha$ distinguishing by triangles in a common network and triangles in a differential network. 
The initial estimated triangles contain more false positive than true positives increasingly with $p$ and $n$. This is corrected by our triangle detection procedure (particularly for common edges), which without losing many true positive edges, reduces notably the number of false positives. 

\begin{table}[h]
\scriptsize
\begin{center}
\caption{{\footnotesize True positives - false positives for the weakest estimated triangle edges such that edge are removed with several significant levels $\alpha$. We also describe the initial estimate, thus not using the triangle correction (NO row).}} 
\begin{tabular}{ l| r r r r  r r r r r}
&\multicolumn{4}{c}{common edges} &&\multicolumn{4}{c}{differential edges} \\

n & $25$ & $100$ &$250$& $500$&& $25$ & $100$ &$250$& $500$\\
\hline
&&\multicolumn{8}{c}{dimension p=200} \\
NO                 &3.53-26.40 & 6.03-40.04&6.39-61.02& 5.75-57.06&&.38-3.68&0.76-4.30&0.51-4.75&0.48-5.10\\
$\alpha=.01$&0-0     &0.09-0.53&1.72-0.97&3.38-1.26&&0-0&0-0.12&0.07-0.48 &0.12-0.76\\
$\alpha=.03$&0-0.02&0.61-1.25&2.93-2.40&4.21-3.21&&0-0&0.03-0.49&0.17-0.94&0.29-1.14\\
$\alpha=.05$&0-0.08&1.21-1.87&3.67-4.30&4.55-5.37&& 0-0&0.09-0.82&0.19-1.27&0.37-1.47\\
&&\multicolumn{8}{c}{dimension p=300} \\
NO                 &5.92-60.20 &9.43-74.25 &8.12-91.25&7.23-114.64&&.51-9.13&0.84-8.35&0.67-7.76&0.38-8.20\\
$\alpha=.01$&0-0&0.30-0.71&2.33-1.09&4.37-1.87&&0-0&0-0.16&0.09-0.67&0.08-0.92\\
$\alpha=.03$&0-0.02&1.03-1.60&3.65-3.89&5.19-5.86&&0-0&0.02-0.65&0.25-1.22&0.21-1.43\\
$\alpha=.05$&0.04-0.10&1.93-3.28&4.52-7.48&5.63-11.11&&0-0.02&0.08-1.05&0.36-1.83&0.28-2.09\\
&&\multicolumn{8}{c}{dimension p=400} \\
NO                & 11.90-232.4 &18.36-241.2&16.43-259.4&13.29-274.3&&.56-17.20&1.14-17.86&0.92-16.69&0.64-17.7\\
$\alpha=.01$&0-0.08&0.7-1.7&4.31-3.36&7.49-5.46&&0-0&0-0.44&0.05-0.98&0.25-1.21\\
$\alpha=.03$&0-0.12&2.09-5.09&6.74-13.22&9.23-19.8&&0-0&0.02-1.26&0.30-1.79&0.40-2.52\\
$\alpha=.05$&0.01-0.37&3.75-12.19&8.30-27.03&9.95-38.5&&0-0&0.14-1.95&0.43-2.85&0.47-4.27\\
\end{tabular}
\label{tabPOWER16}
\end{center}
\end{table}

\section{Network analysis of colon cancer gene expression data}\label{SEC5}
We apply the methods to a real case study. A gene expression dataset which can be downloaded at \url{http://www.ebi.ac.uk/arrayexpress/} and is presented in \cite{Hinoue2012}. A total of 25 patients are examined, the gene expression profiling is obtained in each one of them for a colorectal tumor sample and its healthy adjacent colonic tissue: in total there are 50 samples and  24,526 genes.

\subsection[Reduction of the number of genes to be analysed and clustering]{Reduction of the number of genes to be analysed and \\clustering}\label{prepro}
We reduce the dimension size of the dataset by considering two filters with the objective to select highly correlated genes and differentially correlated genes. For the first filter we use a simple statistic, the adjusted squared correlation, that measures the global strength of gene connections by 
\begin{equation}
\text{AdCor}(g) =  \left(\frac{n-1}{n-2}\right)\frac{\text{SqCor}(g)*p - 1}{p-1} - \frac{1}{n-2},
\end{equation}
with
$$
\text{SqCor}(g) = \frac{1}{p} \sum_{i=1}^p \hat{r}_{ig}^2.
$$
where $\hat{r}_{ig}$ is the sample correlation coefficients for pair of variables $(i,g)$.
We compare the statistic for each gene (independently for the two medical conditions) against a null distribution, $H_0: \{ r_{ig}=0$, $\forall i\neq g\}$. To account for gene dependencies we approximate an empirical null distribution by simulating $n$ independent observations from a normal distribution $N(0,1)$ and then finding the adjusted square correlations between simulated observations and all remaining genes in the real data. 
For the second filter we consider a sum of squares based statistic that uses the differences between Fisher transform healthy and tumor sample correlations:
\begin{equation}\label{sumSquaresTestStatistic}
T_{SS}(g) = \frac{2(n-3)}{p-1} \sum_{j\neq g} [g(\hat{r}_{T_{jg}}) - g(\hat{r}_{H_{jg}})]^2, 
\end{equation}
where $g(\hat{r}_{T_{jg}})$ and $g(\hat{r}_{H_{jg}})$ are the Fisher transform function applied to the sample correlation coefficients for tumor and healthy genes respectively.
For each gene we compare the mean square Fisher transform correlation differences with the expected value under the null hypothesis, say  $H_0: \{(r_T)_{ig}-(r_H)_{ig} =0$, $\forall i \neq g\}$. The null distribution is approximated  using permuted samples. 

In both presented tests we use a  false discovery rate correction \citep{Benjamini1995a} for the p-values to account for multiple testing and a threshold of $0.01$ such that we select genes $g^*$:
$$
g^*= \{g: \text{p-val}(g)^H<0.01\} \cup  \{g: \text{p-val}(g)^T<0.01\} \cup \{g: \text{p-val}(g)^D<0.01\},
$$
where $\text{p-val}(g)^H$ are the adjusted sum of square square test p-values for the healthy dataset,  $\text{p-val}(g)^T$ are the adjusted sum of square square test p-values for the tumor dataset and $\text{p-val}(g)^D$ are the adjusted differential sum of square square test p-values using both tumor and healthy samples. 

\begin{figure}[H]
\begin{center}
\includegraphics[width=8cm,height=5cm]{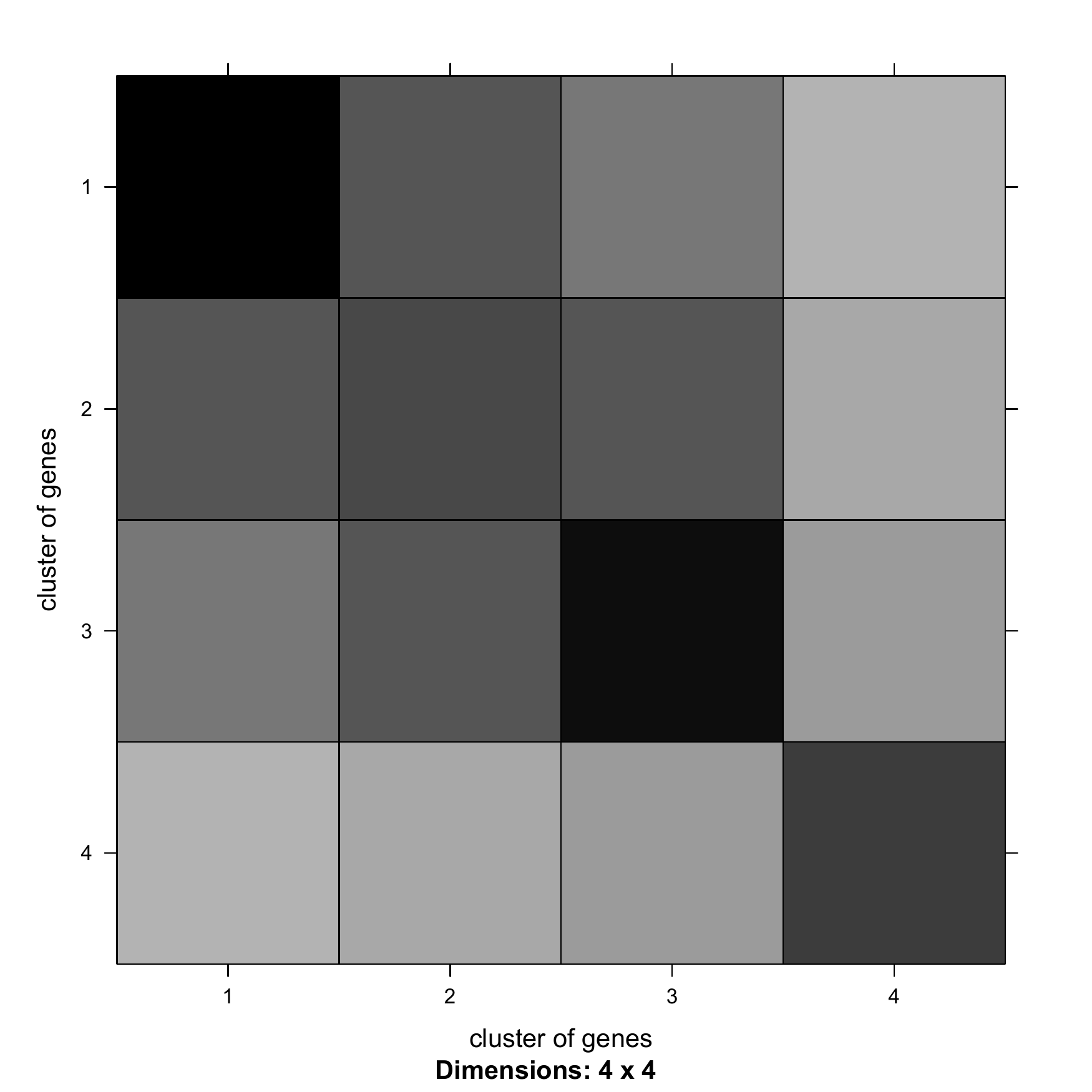} \caption{\footnotesize{Heat-map for between and within gene cluster square correlation averages. Darkness indicates magnitude of the average correlation. }}\label{f41}
\end{center}
\end{figure}

The total length of the reduced genes is $11,163$ which is a reduction of the $54.5\%$ of the variables. We further use a clustering procedure on the reduced dataset to estimate joint networks separately for different groups of genes. We consider the hierarchical clustering algorithm presented in  \cite{Mullner2013} since it provides a  fast procedure even for very large dimensions. We use 1 minus the matrix of correlations for healthy genes as  dissimilarity matrix to find 4 large  clusters of size $[$2582, 4958, 3409, 214$]$ genes respectively. In Figure \ref{f41} we present the heat map of the average square correlation between and within clusters. Note that the darkest squares are given in the diagonal indicating large within cluster correlation magnitudes in comparison to between correlation magnitudes. Moreover, cluster 2 is quite correlated with cluster 1 and 3.
 
\subsection{Network estimation of cancer and healthy gene expression data}\label{SjointReal}
We fit  four weighted fused graphical lasso models corresponding to the 4 clusters of genes defined in Section \ref{prepro}. We use significant levels $\alpha_1$ and $\alpha_2$ to tune the penalization parameters. For $\alpha_1$ we set the underlying expected number of false positive edges (EFP) with EFP = $300, 500, 400, 100$ respectively for each cluster. Then, $\alpha_1 = EFP/ p'$ with $p' = p(p-1)/2$. In terms of $\alpha_2$ we use three different levels which are specified in Table \ref{edges}. 

Precisely, in Table \ref{edges} we show the number of estimated edges common to the two medical conditions and the number of differential edges: healthy for edges only present in the network for healthy samples; and 
tumor for edges  only present in the network for tumor samples. The total number of edges is much larger than the expected number of false positives which suggests certain strength in the results. Moreover, we observe that the number of differential edges is remarkably larger for healthy samples than for tumor samples for all clusters.

\begin{table}[h]
\scriptsize
\begin{center}
\caption{ {\footnotesize Number of edges for common networks and differential edges using several similarity tuning parameters $\alpha_2$.} }  
\begin{tabular}{ l r r r r  r  r r }
&\multicolumn{3}{c}{Cluster 1}&&\multicolumn{3}{c}{Cluster 2}\\
\cline{2-4}\cline{6-8}
$\alpha_2$ &0.001& 0.01 &0.05&&0.001& 0.01&0.05\\
\hline
common&1,386&1,338&1,305&&1,971&1,913&1,927\\
healthy only   &26& 70&518&&53&220&958\\
tumor only     &1&2&20&&3&22&116\\
\hline\\
&\multicolumn{3}{c}{Cluster 3}&&\multicolumn{3}{c}{Cluster 4}\\
\cline{2-4}\cline{6-8}
$\alpha_2$& 0.001& 0.01& 0.05&&0.01& 0.03&0.05\\
\hline
common& 2,542&2,315&2,024&&124&123&116\\
healthy only    	&104 &355&1,129&&0&1&2\\
tumor only     &5 &28&111&&0&0&0\\
\end{tabular}
\label{edges}
\end{center}
\end{table}
 
Differential network uncertainty is assessed by applying the permutation process proposed in Section \ref{uncertainty} which estimates WFGL for data under the hypothesis of $\Omega_X=\Omega_Y$. We use  $20$ iterations for the four clusters of genes (and only using the second value of $\alpha_2$) to get a glimpse of the expected networks sizes under the assumption of equality between the two networks. In cluster 1, the estimated differential edges for either healthy or tumor samples using permuted samples range from $0$ to $30$ with an average of about $14$ edges. The estimated number of only healthy edges is much larger with $70$ edges, whereas only tumor edges ($2$) are exceeded by the $95\%$ of the permuted data estimates. In cluster 3, we expected between $25$ to $129$ differential edges under the null hypothesis whereas for only healthy samples we observe a total of $355$ edges. Finally, in cluster 4, estimated differential edges are within the range observed for permuted samples.

\begin{figure}[h]
\begin{center}
 \begin{tabular}{cc}
\subfloat[Network estimation in cluster 1]{\includegraphics[scale=0.35]{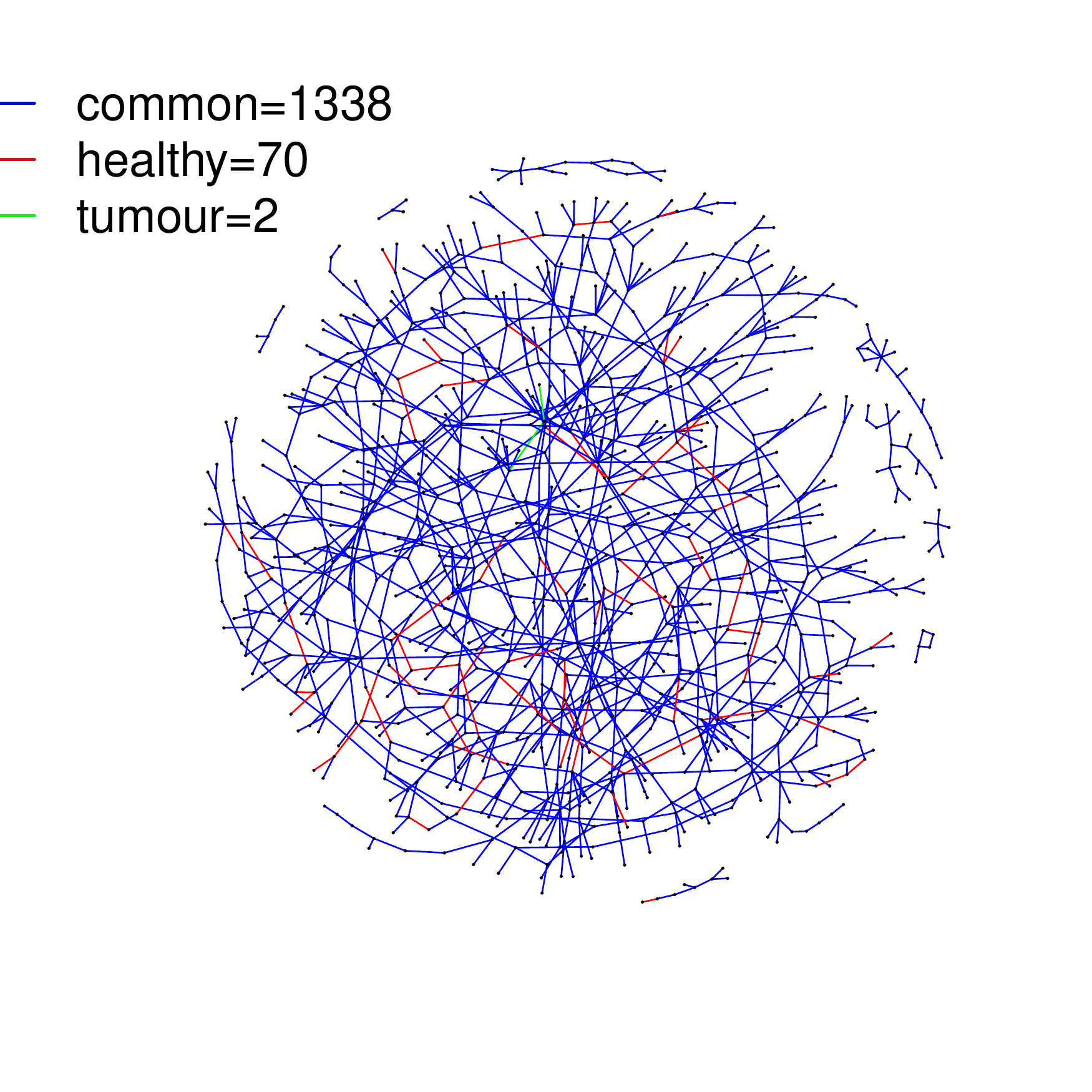}}&
\subfloat[Network estimation in cluster 2]{\includegraphics[scale=0.35]{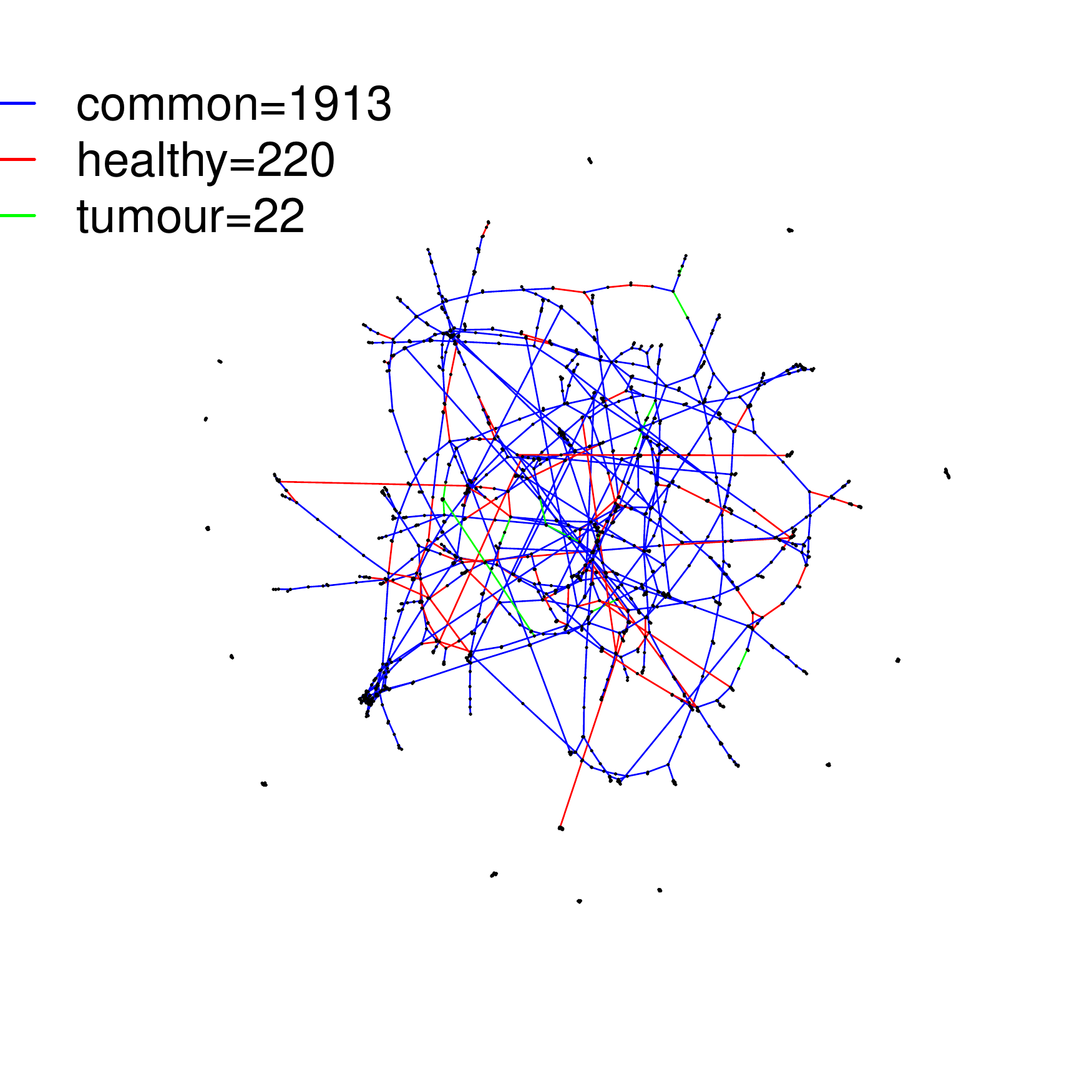}}\\
\subfloat[Network estimation in cluster 3]{\includegraphics[scale=0.35]{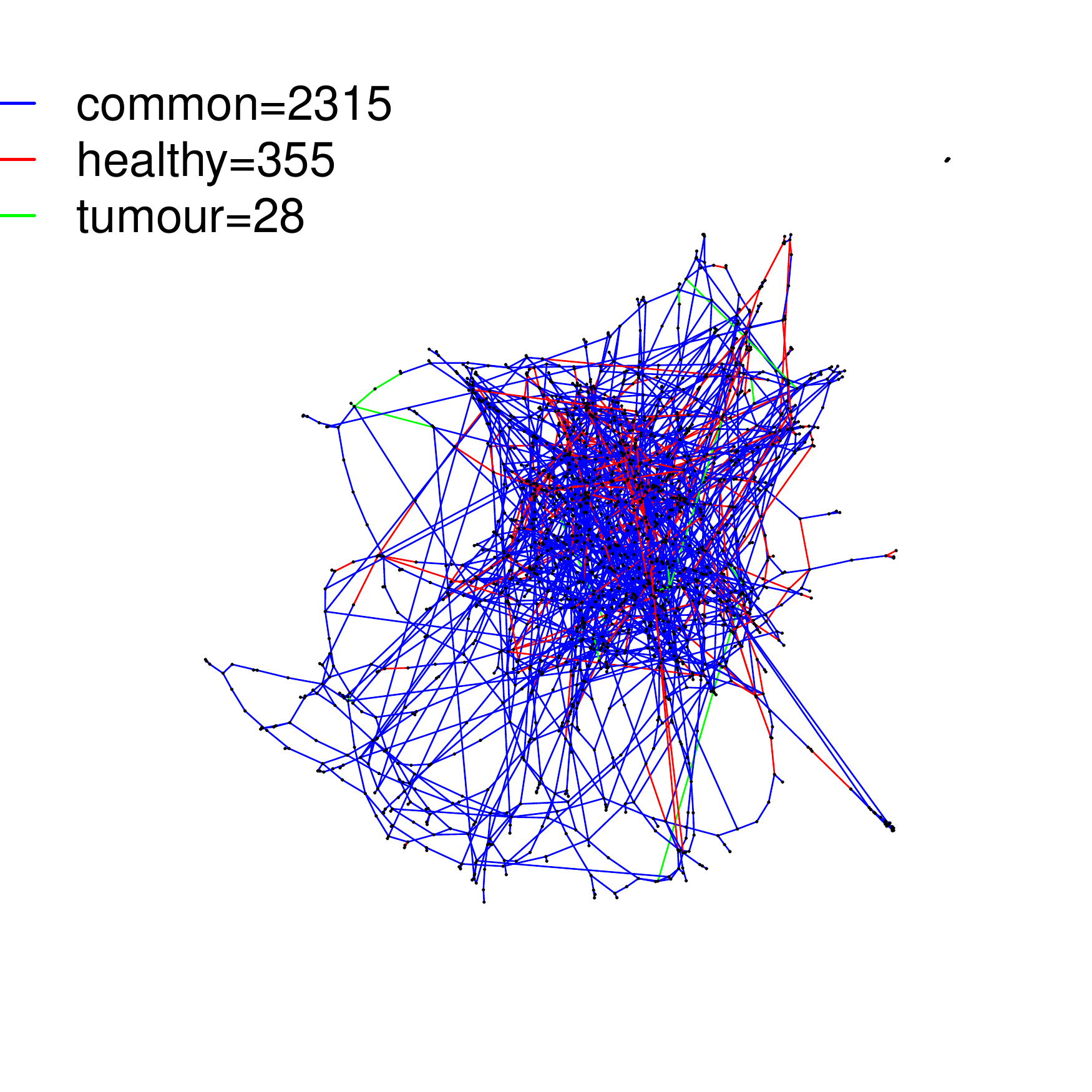}}&
\subfloat[Network estimation in cluster 4]{\includegraphics[scale=0.35]{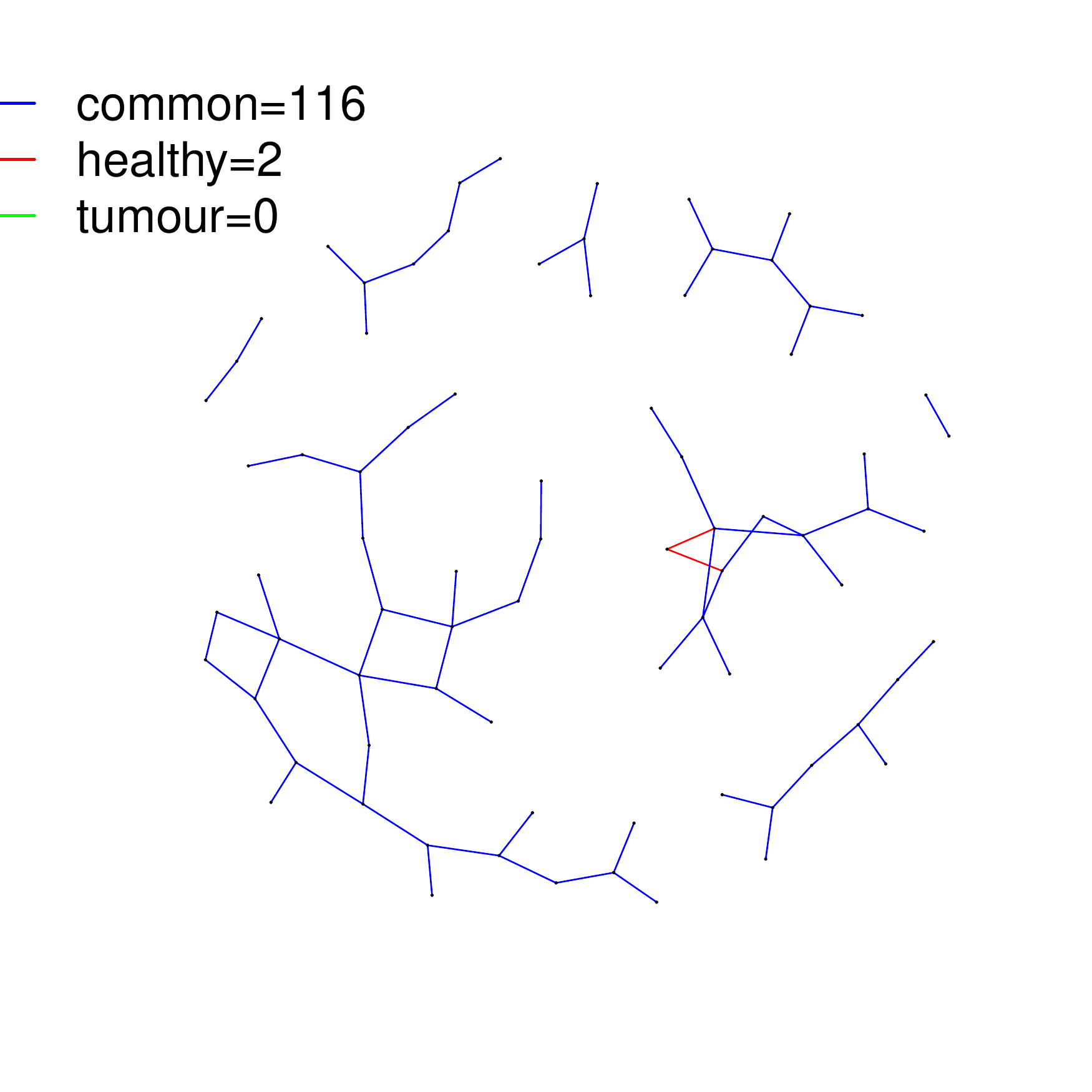}}
\end{tabular}
\caption{\footnotesize{Estimated joint networks for four groups of genes: in blue there are the common edges and in green (healthy) and red (tumour) the differential connections.}}\label{f44}
\end{center}
\end{figure}

In Figure \ref{f44} we show the graphical representation of some of the estimated networks. The blue edges are  common edges, whereas in red there are only healthy edges and in green there are tumor edges.  In general, in almost all clusters we detect presence of hub genes (genes with much higher degree than the rest). Furthermore, we can see a clustered graph structure in the estimated networks, which could be expected in biological data \citep{Eisen1998} with some specific groups of genes that are only present in one medical condition (especially healthy genes).

\subsection{Integration with biological pathway lists}\label{SjointPath}
We are particularly interested in knowing how standard gene pathways change in different medical conditions. To assess which biological processes might be linked to changes in the gene connections we download $1,320$ gene sets from the MSig database (http://www.broad institute.org/gsea/msigdb/index.jsp), which represent canonical pathways compiled from two sources: KeGG (\url{http://www.genome.jp/kegg/pathway.html}) and Reactome (\url{http://www.reactome.org/}). 

Note that in the original data some genes are represented by more than one probe  (these are not identical, so they are not merely technical replicates), which we have considered as separate variables in the graph estimation. Overall 11,163 such variables are used after the dimension reduction corresponding to 9,571 different genes. In the joint estimation stage we have estimated 436 connections (using the largest $\alpha_2$) that correspond to pairs of variables describing the same gene in different proves. This nears the $23\%$ of the total possible connections in this setting. For all following summaries and analyses in this section we give results on gene level rather than on probe level in order not to inflate results with spurious correlations.

To integrate and analyze the estimated networks within the pathway lists, we count which pairs of connected genes in the estimated networks are both present in a specific pathway list (see Table \ref{pathConn}). Using the 9,571 genes as background, we find that approximately $1\%$ of estimated connections are expected to be included by chance. Thus, in the table we also evaluate how likely it is to obtain at least the same  number of biological relevant connections in a random process (given between brackets by an empirical p-value). Common network associations are significantly present in pathways for all clusters. Moreover, tumor networks, which as seen in Table \ref{edges} do not have many singular connections, have significant overlap with the pathways in the third cluster. Healthy connections are rarely significant in any of the four clusters (only cluster 1 shows a significant value). 

\begin{table}[ht]
\scriptsize
\begin{center}
\caption{ {\footnotesize Total number of estimated edges whose genes are both in the same pathway list (p-value}. }  
\begin{tabular}{ l r r r r  r  r r }
&\multicolumn{3}{c}{Cluster 1}&&\multicolumn{3}{c}{Cluster 2}\\
\cline{2-4}\cline{6-8}
$\alpha_2$ &0.001& 0.01 &0.05&&0.001& 0.01&0.05\\
\hline
common& 48(0.000)&   47(0.000) &   46(0.000)&& 70(0.000) &   69(0.000) &   74(0.000)\\
healthy only   &0 (1.000) &2 (0.164) &12 (0.012)&& 1(0.439)& 1(0.901)& 8 (0.823)\\
tumor only      &0 (1.000)& 0 (1.000)& 0 (1.000)&&0 (1.000)& 1 (0.218)& 2 (0.367) \\
\hline\\
&\multicolumn{3}{c}{Cluster 3}&&\multicolumn{3}{c}{Cluster 4}\\
\cline{2-4}\cline{6-8}
$\alpha_2$& 0.001& 0.01& 0.05&&0.01& 0.03&0.05\\
\hline
common&74 (0.000)&   70 (0.000)&   60 (0.000)&&3 (0.109)&   3 (0.107)&   3 (0.095)\\
healthy only &2 (0.306)& 6 (0.195)& 18 (0.117)&& 0 (1.000)& 0 (1.000)& 0 (1.000)\\
tumor only     &0 (1.000)& 0 (1.000)& 5 (0.008)&&0 (1.000)& 0 (1.000)& 0 (1.000)\\
\end{tabular}
\label{pathConn}
\end{center}
\end{table}

The most frequent pathway for only healthy connections is {\it reactome immune system} with a total of 5 appearances. 
In the only tumor associations, the pathways {\it reactome signaling by GPCR} and {\it reactome GPCR downstream signaling} have both 3 appearances and have been associated with cancer in recent studies \citep{Dorsam2007}. In the common network, several pathways are involved. For instance  {\it reactome cell cycle}, {\it reactome cell cycle mitotic} and {\it reactome immune system} occur in 32, 26 and 34 connections.

\begin{figure}[h]
\begin{center}
 \begin{tabular}{cc}
\subfloat[Immune system genes]{\includegraphics[scale=0.35]{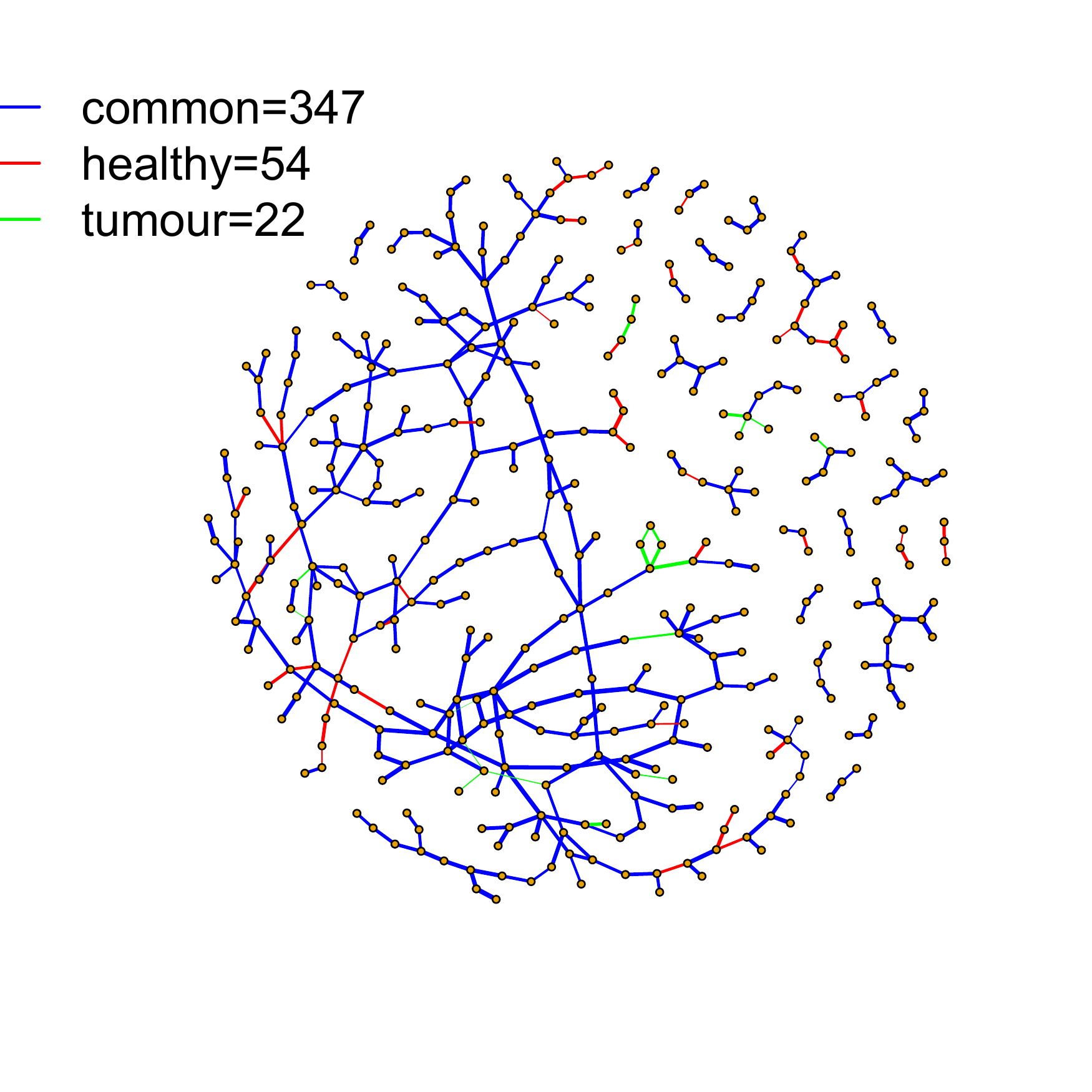}}&
\subfloat[GPCR genes]{\includegraphics[scale=0.35]{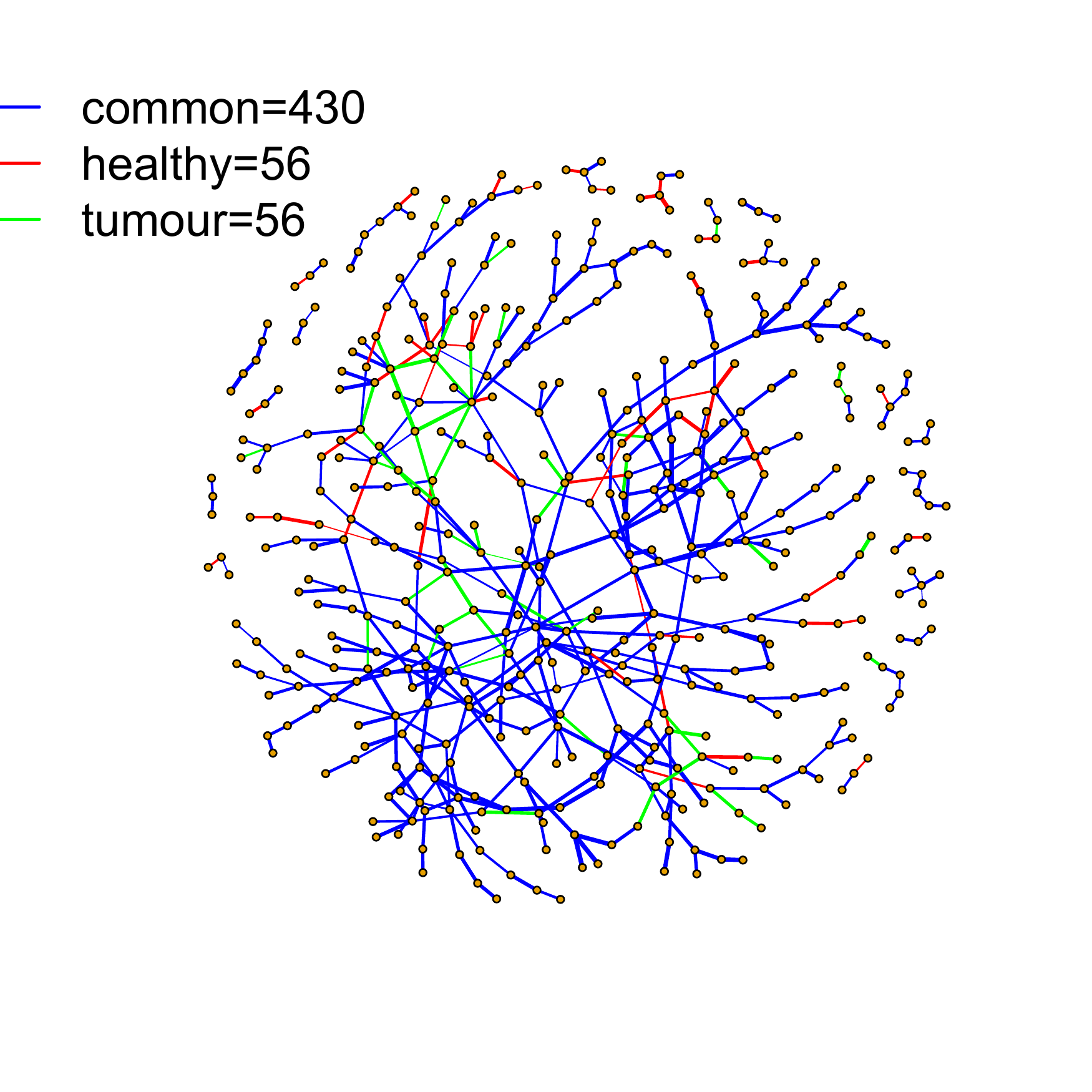}}\\
\end{tabular}
\caption{\footnotesize{Estimated joint networks for genes in two different patways: {\it reactome immune system} and {\it reactome signaling by GPCR}.}}\label{f49}
\end{center}
\end{figure}

We perform further investigation for genes in the pathways  {\it reactome immune system} and {\it reactome signaling by GPCR}. We estimate a joint conditional dependence structures only considering the genes in each of the two pathways. In Figure \ref{f49} we show their  graphical representation using $\alpha_1=0.001$ and $\alpha_2=0.05$ which supposes an expected false positive edges of about $330$ in both lists. The thickness of the edges indicate the strength of the connection using the average square correlation coefficient (or the square correlation difference between the two conditions for differential edges). For immune system genes, the $70\%$ of the estimated differential edges are due to destroyed connections in tumor samples whereas a 50/50 differential edges relationship exists for GPCR.
 
Differential network uncertainty is also assessed by applying the permutation process proposed in Section \ref{uncertainty}. We use  $100$ iterations for both immune system and GPCR pathway lists. For the immune system, the number of only healthy edges is not expected by chance (with non of the permuted sample estimations exceeding the $54$ edges). In contrast, the number of only tumor edges of $22$ is exceeded by the $28\%$ of the replicates. The sum of differential edges (tumor plus healthy) is also highly significant with the observed maximum out of the $100$ repetitions only reaching $59$ edges (for the $76$ we have estimated in total). For the GPCR pathway, the number of only healthy edges and also only tumor edges is exceeded for permuted-based estimations in only the $2\%$ of the cases and the sum of the two type of differential edges of $112$ is far away from the observed maximum of $89$ edges under permuted samples. This reinforce the power of the results given that the real dataset sample size is very small in comparison to the the dimension.

\section{Discussion}
In this article we present a method to estimate a joint precision matrix in high-dimensional dependent datasets. As for the approach proposed by \cite{Danaher2014}, we consider a penalized maximum likelihood estimator that assumes both sparsity and similarity between the two conditional dependence structures corresponding to the two populations. 

Motivated for our application to genomic data in which gene expression is obtained for the same individual in two different conditions, we develop methodology to account for dependent data. We argue that this consists in a fairer procedure in the sense that all differential connections with same true partial correlation coefficients have approximately the same chance to be recovered. This is confirmed in our simulated data study, where we identify that estimated partial correlation coefficients in the two populations are correlated and that it has to be taken into account to construct fair differential networks. 

A method to select the tuning parameters in the joint graphical lasso algorithm is also presented in this paper. We monitor the expected false positive rate (EFPR) in order to select the hyper-parameters. We use robust statistics for variance estimators to transform the selection problem to the more intuitive selection of expected proportion of false positive edges. Then, we see in the simulated data analysis that the proposed method produces results near the desired EFPR for a sufficiently large sample size.

Finally we detail the problems of the recursive algorithm to estimate closed-form graph structures as triangle of variables. We present a method to correct for this issue  that assesses the evidence of the weakest edge in a triangle structure after the estimation process. Using simulated data we corroborate that our proposed strategy reduces the number of false positive edges without missing many estimated true positives.

The analysis of the motivating gene expression data with healthy and tumor populations underlines some interesting results. We estimate 4 joint networks corresponding to 4 clusters of genes. In all 4 estimated networks we see how genes interact between each other in groups, suggesting a clustering structure.  In all of them, healthy gene connections are more present than tumor connections.  Further pathway integration analysis suggest that common edges, which are estimated using a larger effective sample size than the 25 individuals, have a strong significant overlap with these pathway lists.  Moreover, genes in the immune system suffer the general behavior in the differential network (described by more only healthy than only tumor edges) whereas genes in the GPCR have similar number of  differential edges in the two medical conditions.




\normalsize
\section{Supplementary material}
\subsection{Estimation of dependence structure measure}\label{psiEST}
In Section \ref{pairedWeights} we proposed two different ways to estimate the correlation between same coefficients in $\hat{\Omega}_X$ and $\hat{\Omega}_Y$ for similarity penalization. Here we analyze the performance of the estimators using simulated data. We calculate the mean square error of $[\hat{\psi}_{ij}]$ against $[\psi_{ij}]$  as well as the correlation $\text{cor}(\psi,\hat{\psi})$. We compare the  Reg-based (eq. \ref{icorSepEst}) and Reg-based-sim (eq. \ref{icorSepEst2}) estimator results with $[\hat{\psi}_{ij}=0.5]$ (which assumes independence between samples). The values $\psi_{ij}$ are approximated by the sample correlation using $5,000$ iid Monte Carlo replicates of the theoretical model. In Table \ref{tabPOWER11} we present the average ranks  (average MSE) for the mean square error and in Table \ref{tabPOWER12} we give the average ranks (average correlation) for the correlation levels. Rank = 1 is assigned to the best estimator and Rank = 3 is given to the worst estimator. 

For very small sample sizes ($n = 25$), the estimators' MSE are very large, and can even find worse results than assuming independence. However, for all other investigated sample sizes, the Reg-based and its simplified version find the lowest MSE. Correlation-wise, the two proposed estimators give large positive correlations consistently for large $p/n$ ratios.

\begin{table}[h]
\scriptsize
\begin{center}
\caption{\footnotesize{Ranks and average for the sum of MSE.}} 
\begin{tabular}{ l r r r r r}
n & $25$ & $50$ &$150$& $300$ & $500$\\
\hline
\multicolumn{6}{c}{dimension p=50} \\
Reg-based     & 2.04 (0.86) & 1.83 (0.42)& 1.52 (0.16)& \textbf{1.40 (0.09)} &\textbf{1.28 (0.06)}\\
Reg-based-sim&\textbf{1.04 (0.86)}& \textbf{1.17 (0.41)}& \textbf{1.48 (0.16)}& 1.60 (0.09)& 1.72 (0.06)\\
Independence  &2.91 (1.24) & 3.00 (1.30) & 3.00 (1.38) & 3.0 (1.41) & 3.00 (1.42)\\
\multicolumn{6}{c}{dimension p=170} \\
Reg-based    &2.74 (0.74) &  2 (0.33) &\textbf{1.17 (0.13)} & \textbf{1.01 (0.08)} &\textbf{1.06 (0.06)}\\
Reg-based-sim&1.74 (0.74)&   \textbf{1.00 (0.33)}& 1.83 (0.13)& 1.99 (0.08) &1.94 (0.06)\\
Independence &\textbf{1.52 (0.77)}& 3.00 (0.70) &3.00 (0.65) &3.00 (0.67) &3.00 (0.69)\\
\multicolumn{6}{c}{dimension p=290} \\
Reg-based    & 2.30 (0.74)   & 2.00  (0.33) &\textbf{1.00 (0.13)}  & \textbf{1.00 (0.08)}   &\textbf{1.00 (0.06)}\\
Reg-based-sim& \textbf{1.30  (0.74)} &\textbf{1.00 (0.33)} &  2.00 (0.13)  & 2.00 (0.08)  & 2.00 (0.06)\\
Independence &2.40 (0.77)  & 3.00  (0.70)  &3.00 (0.65)  & 3.00 (0.67)  & 3.00 (0.69)\\
\multicolumn{6}{c}{dimension p=500} \\
Reg-based    &2.80 (0.72)   &2.00 (0.32)    &\textbf{1.00 (0.13)}   &\textbf{1.00 (0.09)} &\textbf{1.00 (0.06)}\\
Reg-based-sim&1.79 (0.72)   &\textbf{1.00  (0.32)}  &2.00 (0.13)  & 2.00 (0.09)& 2.00 (0.06)\\
Independence &\textbf{1.42 (0.68)}   &3.00 (0.64)  & 3.00 (0.58)  & 3.00 (0.59)  &3.00 (0.61)\\
\end{tabular}
\label{tabPOWER11}
\end{center}
\end{table}

\begin{table}[h]
\scriptsize
\begin{center}
\caption{\footnotesize{Ranks and average for the average correlations between approximated and estimated $\psi$.}} 
\begin{tabular}{ l r r r r  r  r r r r}
n & $25$ & $50$ &$150$& $300$ & $500$\\
\hline
\multicolumn{6}{c}{dimension p=50} \\
Reg-based    &\textbf{1.16 (0.63)}& \textbf{1.23 (0.80)}&  \textbf{1.5 (0.93)} & 1.67 (0.96) &1.57 (0.97)\\
Reg-based-sim&1.84 (0.63)&1.77  (0.80)&\textbf{1.5 (0.93)} &\textbf{1.33 (0.96)} &\textbf{ 1.43 (0.97)}\\
Independence &3.00 (0)& 3.00 (0)& 3.00 (0)& 3.00 (0)& 3.00 (0)\\
\multicolumn{6}{c}{dimension p=170} \\
Reg-based    &\textbf{1.04 (0.57)} &\textbf{1.09 (0.69)} &\textbf{1.34 (0.86)} &1.55 (0.92)& 1.94 (0.95)\\
Reg-based-sim&1.96 (0.57)& 1.90 (0.69)&1.66 (0.86) &\textbf{1.45 (0.92)}& \textbf{1.05 (0.95)}\\
Independence &3.00 (0)& 3.00 (0)& 3.00 (0)& 3.00 (0)& 3.00 (0)\\
\multicolumn{6}{c}{dimension p=290} \\
Reg-based    &\textbf{1.07 (0.62)} &\textbf{1.03 (0.72)}& 1.51 (0.86)&  1.90 (0.92) &1.94 (0.95)\\
Reg-based-sim&1.92 (0.62)&1.97 (0.72) &\textbf{1.49 (0.86)}& \textbf{1.10 (0.92)}&\textbf{1.05 (0.95)}\\
Independence &3.00 (0)& 3.00 (0)& 3.00 (0)& 3.00 (0)& 3.00 (0)\\
\multicolumn{6}{c}{dimension p=500} \\
Reg-based    &\textbf{1.28 (0.61)} & \textbf{1.08 (0.73)} &\textbf{1.06 (0.85)} &\textbf{1.16 (0.91)}& \textbf{1.47 (0.94)}\\
Reg-based-sim&1.72 (0.61)&1.92 (0.73) &1.94 (0.85) &1.84 (0.91) & 1.53 (0.94)\\ 
Independence &3.00 (0)& 3.00 (0)& 3.00 (0)& 3.00 (0)& 3.00 (0)\\
\end{tabular}
\label{tabPOWER12}
\end{center}
\end{table}

\subsection[Tuning parameter selection: normality assumption, variance estimator and similarity regularization parameter]{Tuning parameter selection: normality assumption, \\variance estimator and similarity regularization parameter}
\begin{figure}[h]
\begin{center}
 \begin{tabular}{ccc}
    \subfloat[n=25,  differential]{\includegraphics[scale=0.2]{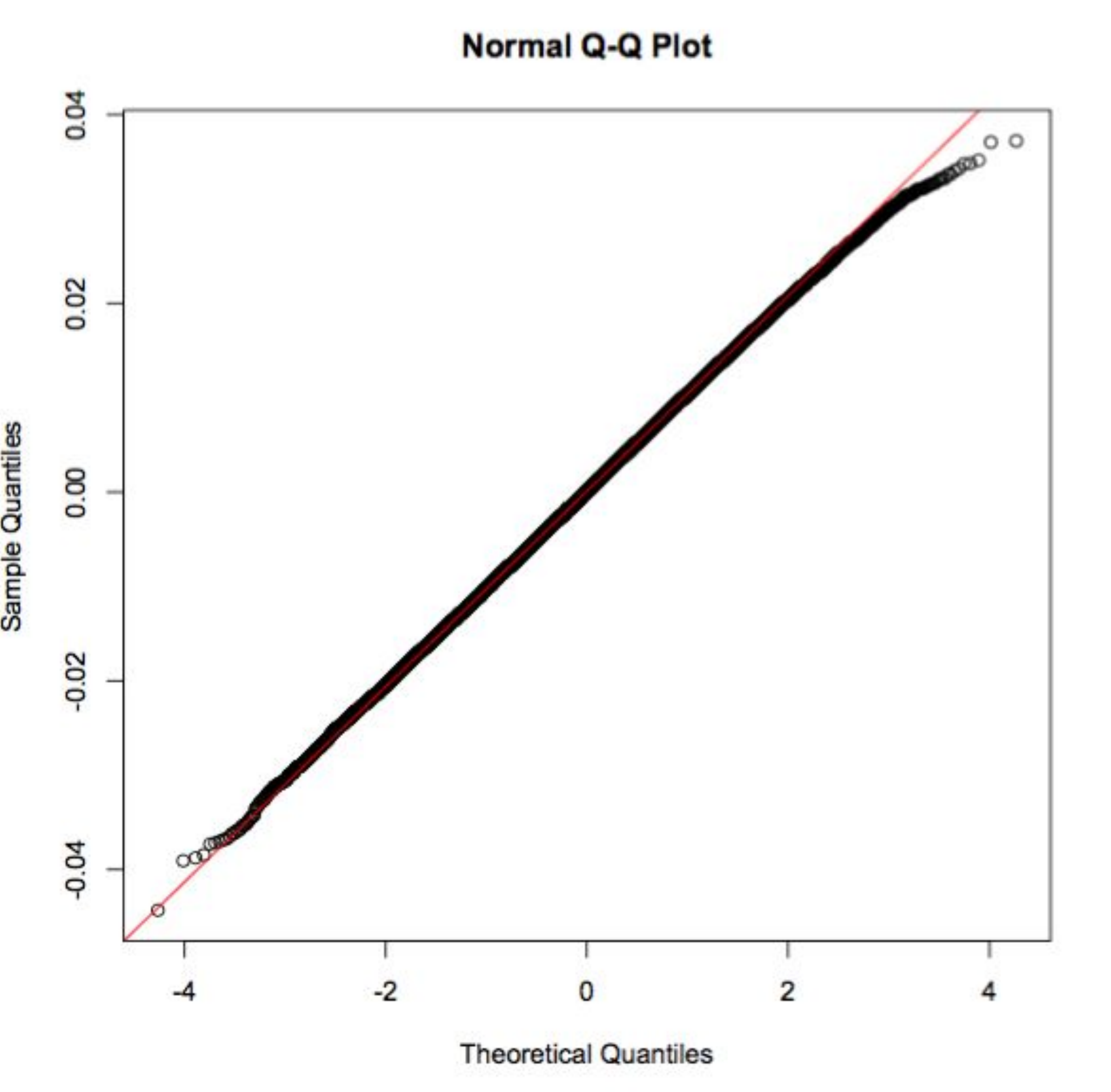}}&
    \subfloat[n=100, differential]{\includegraphics[scale=0.185]{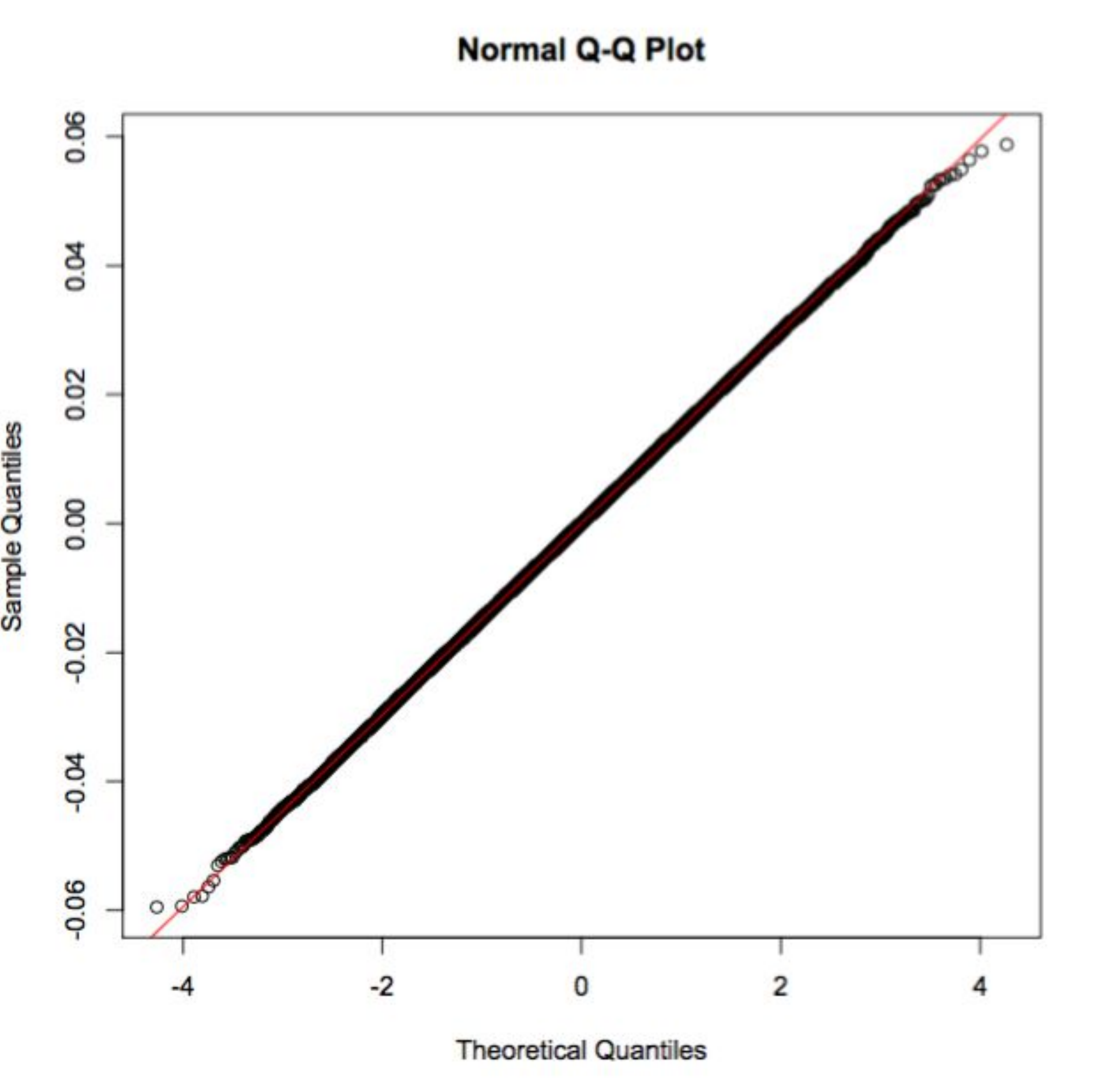}}&
    \subfloat[n=200, differential]{\includegraphics[scale=0.185]{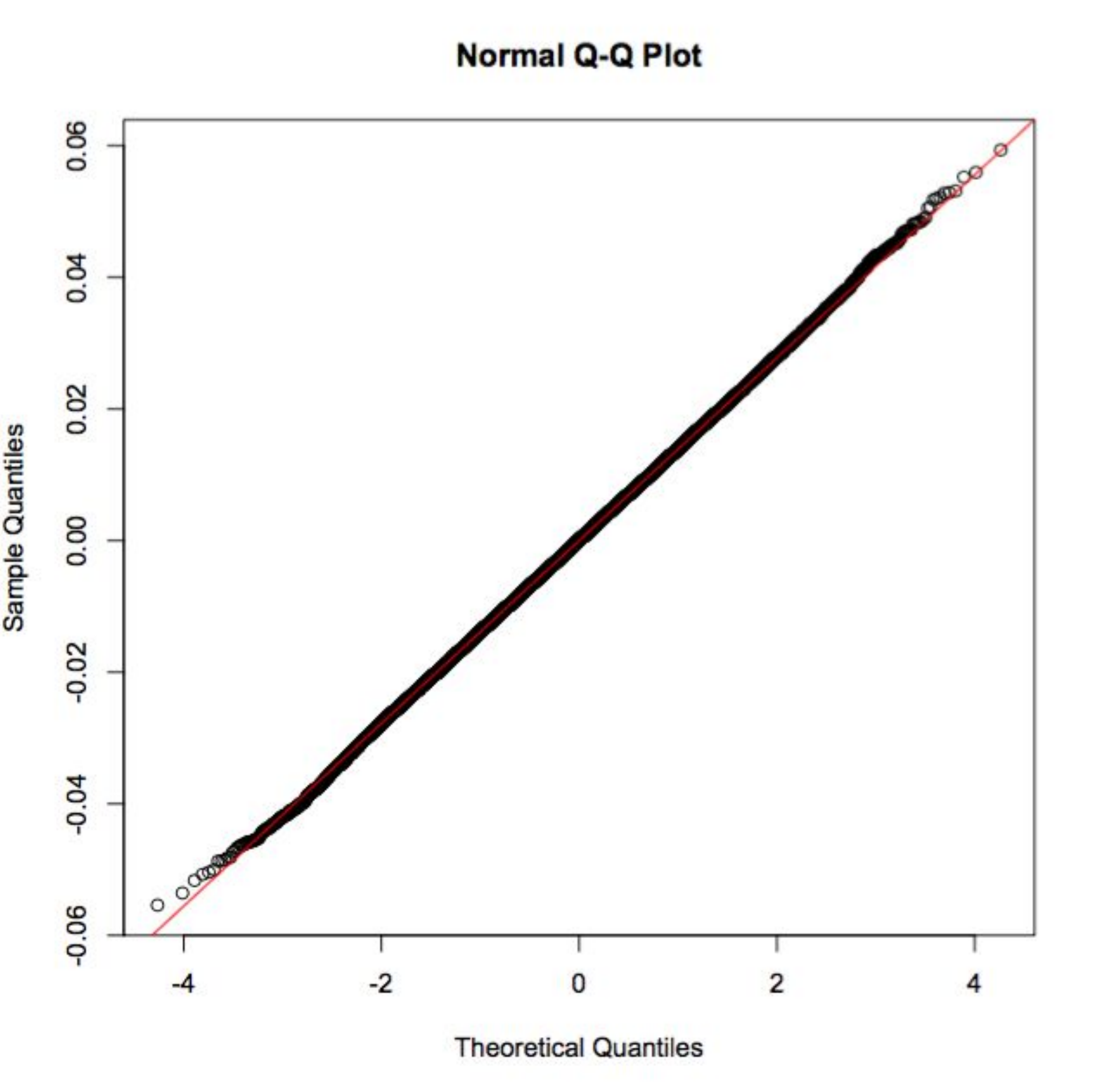}}\\
    \subfloat[n=25, joint]{\includegraphics[scale=0.22]{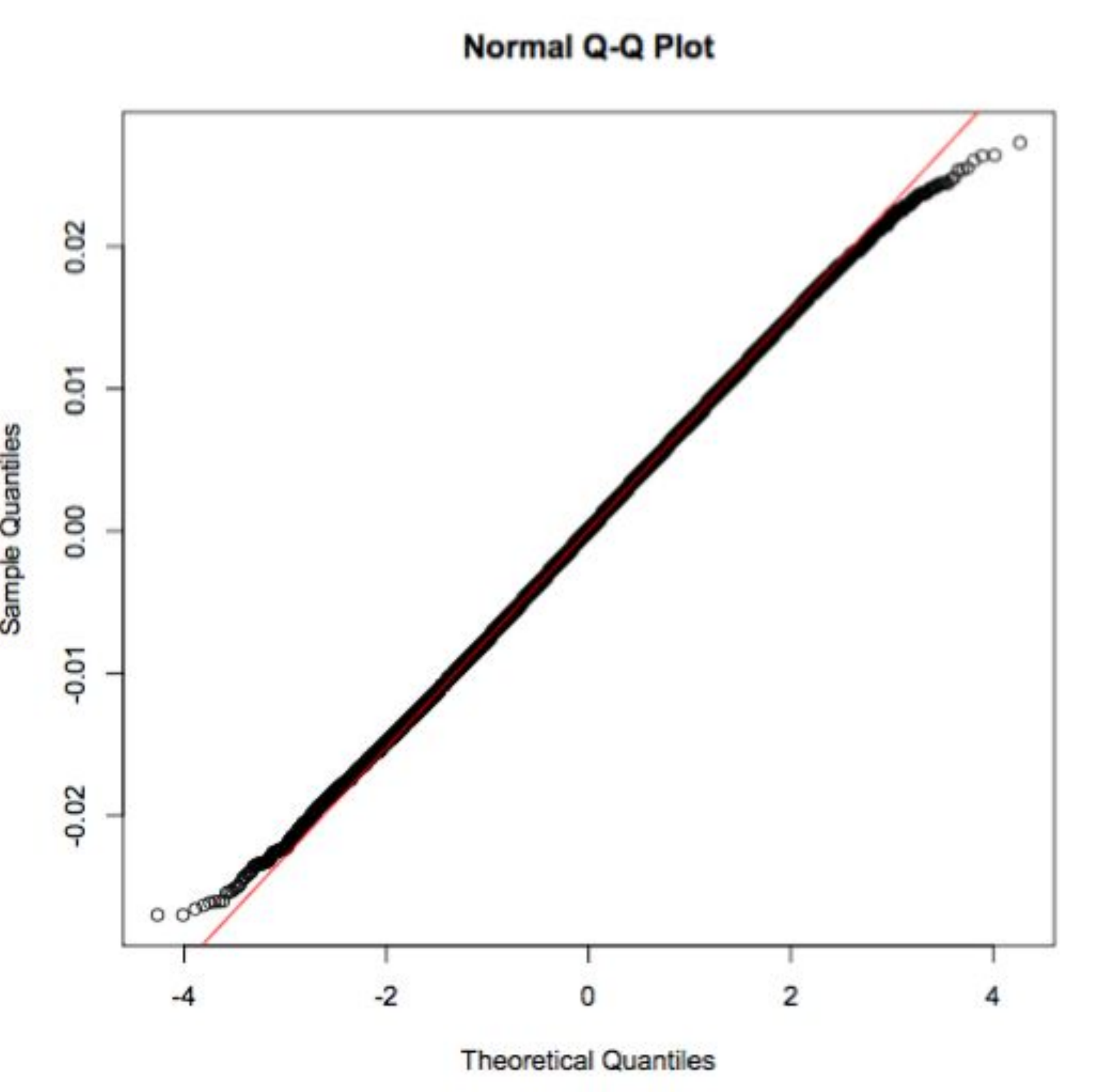}}&
    \subfloat[n=100, joint]{\includegraphics[scale=0.2]{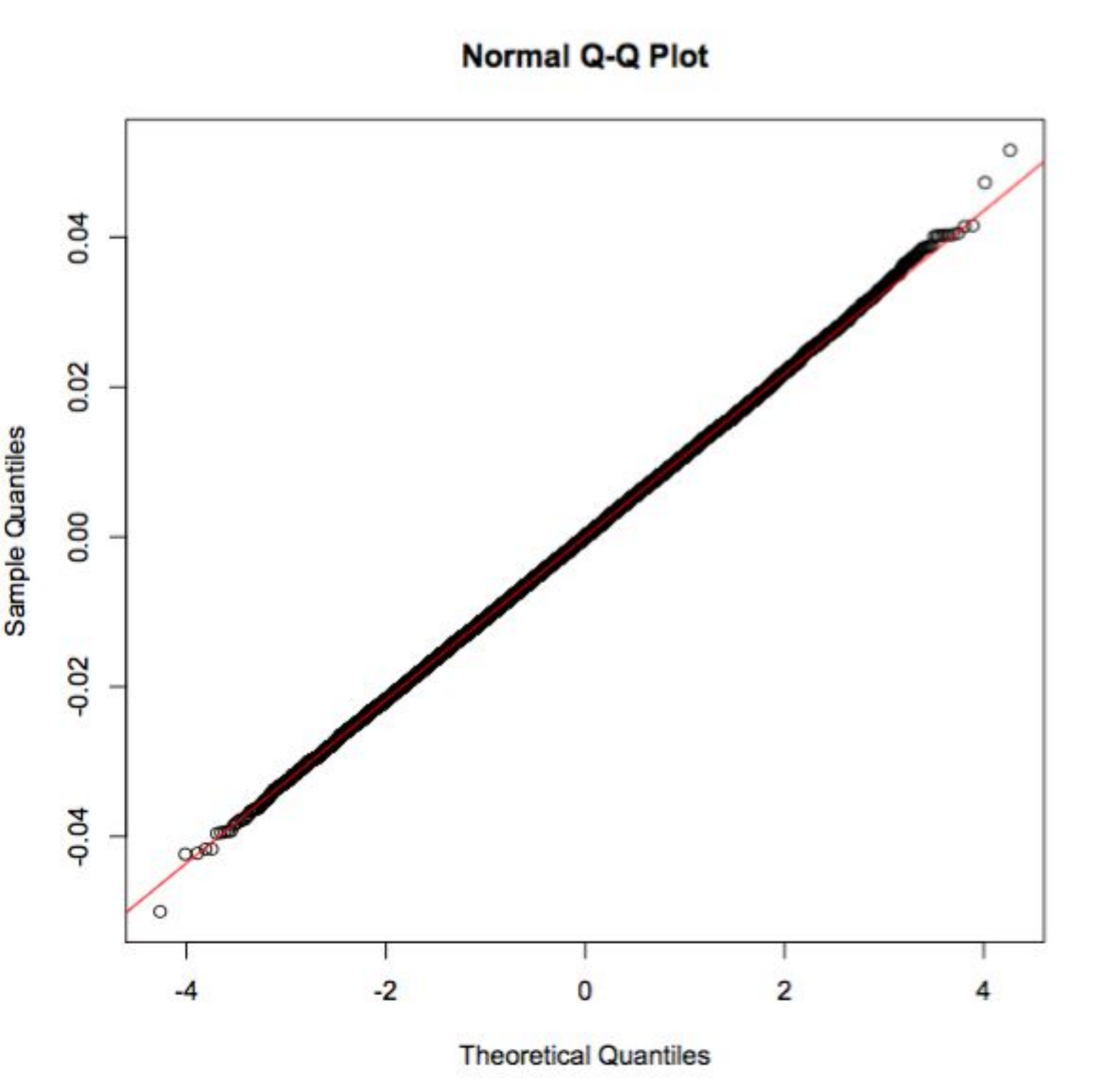}}&
    \subfloat[n=200, joint]{\includegraphics[scale=0.175]{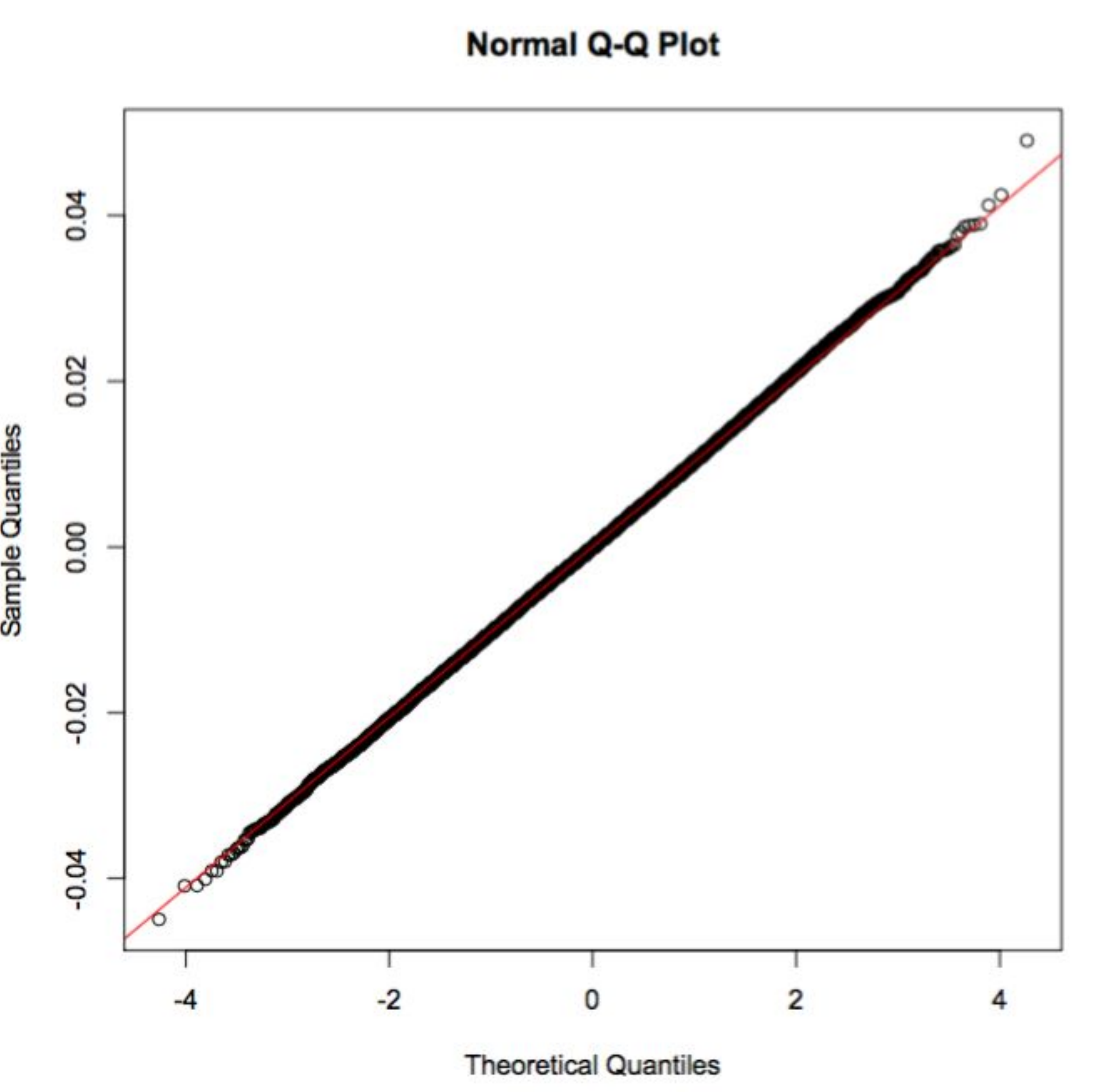}}
     \end{tabular}
 \caption{{\footnotesize qqnorm plots for several examples of estimated precision matrices coefficients. We distinguish between estimated coefficients in both populations (joint) and differential coefficient estimates(differential) using three sample sizes $n$. }}
\label{fSigg6}
\end{center}
\end{figure}

In section \ref{tuning} we discussed a way to select the regularization parameters $\lambda$'s based on setting their correspondent significance levels $\alpha_1$ and $\alpha_2$. We make an assumption of normality for the estimated precision matrix coefficients in each iteration of the joint estimation algorithm. In Figure \ref{fSigg6} we show some of the obtained normality qqplots employing the estimated coefficients as well as the estimated differential coefficients on generated datasets with $p = 300$ and $n = 25, 100, 200$. This represents a general observed behavior in many tested datasets. We shall see that for sufficiently large $n$ the Gaussian assumption is well justified.

The key step in the procedure presented in section \ref{tuning} is the estimation of a robust variance to determine $\lambda$ by the $\alpha$-quantile of a normal distribution. In Table \ref{tabPOWER14}  we compare the performance of three of the most common robust variance estimators: mad, IQR and RCmad.  We show the mean square errors of $\hat{\sigma}_2$ against an approximated $\sigma_2$ found by only using sample partial correlation coefficients whose true values are zero. We compare the performance of the three estimators by their average ranks with rank = 1 being the minimum MSE, and rank = 3 being the maximum MSE. We also provide the average value for 50 simulated datasets. The RCmad estimator finds the best rates as $p$ increases for any sample size $n$ and will be the default approach.

\begin{table}[H]
\scriptsize
\begin{center}
\caption{\footnotesize{Ranks (and average MSE) for the mean square error between approximated and estimated $\sigma_2$.}} 
\begin{tabular}{ l r r r r }
n & $25$ & $100$ &$250$& $500$\\
\hline
\multicolumn{5}{c}{dimension p=200} \\
mad         &\textbf{1.72 (0.93)}& 2.62 (0.92)& 2.66 (1.14)& 2.58 (1.09)\\
IQR          &2.42 (1.29)& 1.78 (0.79)&1.66 (0.66)& 1.86 (0.79)\\
RCmad    &1.86 (0.96)& \textbf{1.60 (0.62)}&\textbf{1.68 (0.53)}& \textbf{1.56 (0.55)}\\
\multicolumn{5}{c}{dimension p=300} \\
mad         &\textbf{1.66 (0.57)}& 2.82 (0.69)& 2.72 (0.67)& 2.38 (0.68)\\
IQR          &2.68 (0.69)& \textbf{1.44 (0.40)}& 1.68 (0.48)& 1.92 (0.63)\\
RCmad    &1.66 (0.61)& 1.74 (0.66)& \textbf{1.60 (0.35)}& \textbf{1.70 (0.47)}\\
\multicolumn{5}{c}{dimension p=400} \\
mad         &1.70 (0.27) &2.64 (0.30)&2.62 (0.36)& 2.68 (0.38)\\
IQR          &2.78 (0.58) &1.80 (0.22)&1.82 (0.24)& 1.74 (0.27)\\
RCmad    &\textbf{1.52 (0.27)} &\textbf{1.56 (0.18)}&\textbf{1.56 (0.16)}& \textbf{1.58 (0.19)}\\

\end{tabular}
\label{tabPOWER14}
\end{center}
\end{table}

\begin{figure}[ht]
\begin{center}
 \begin{tabular}{cc}
    \subfloat[n=25]{\includegraphics[scale=0.28]{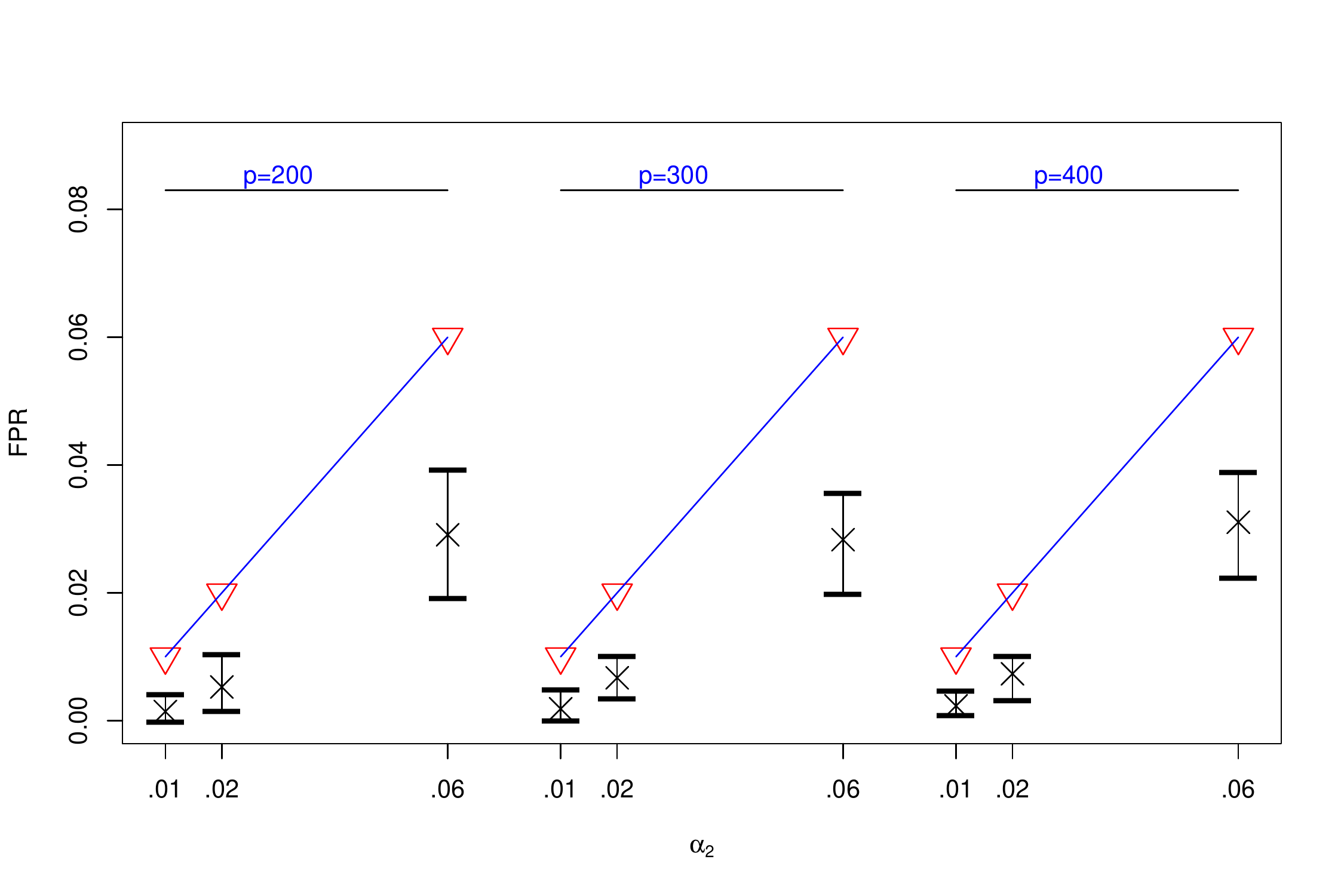}}&
    \subfloat[n=100]{\includegraphics[scale=0.28]{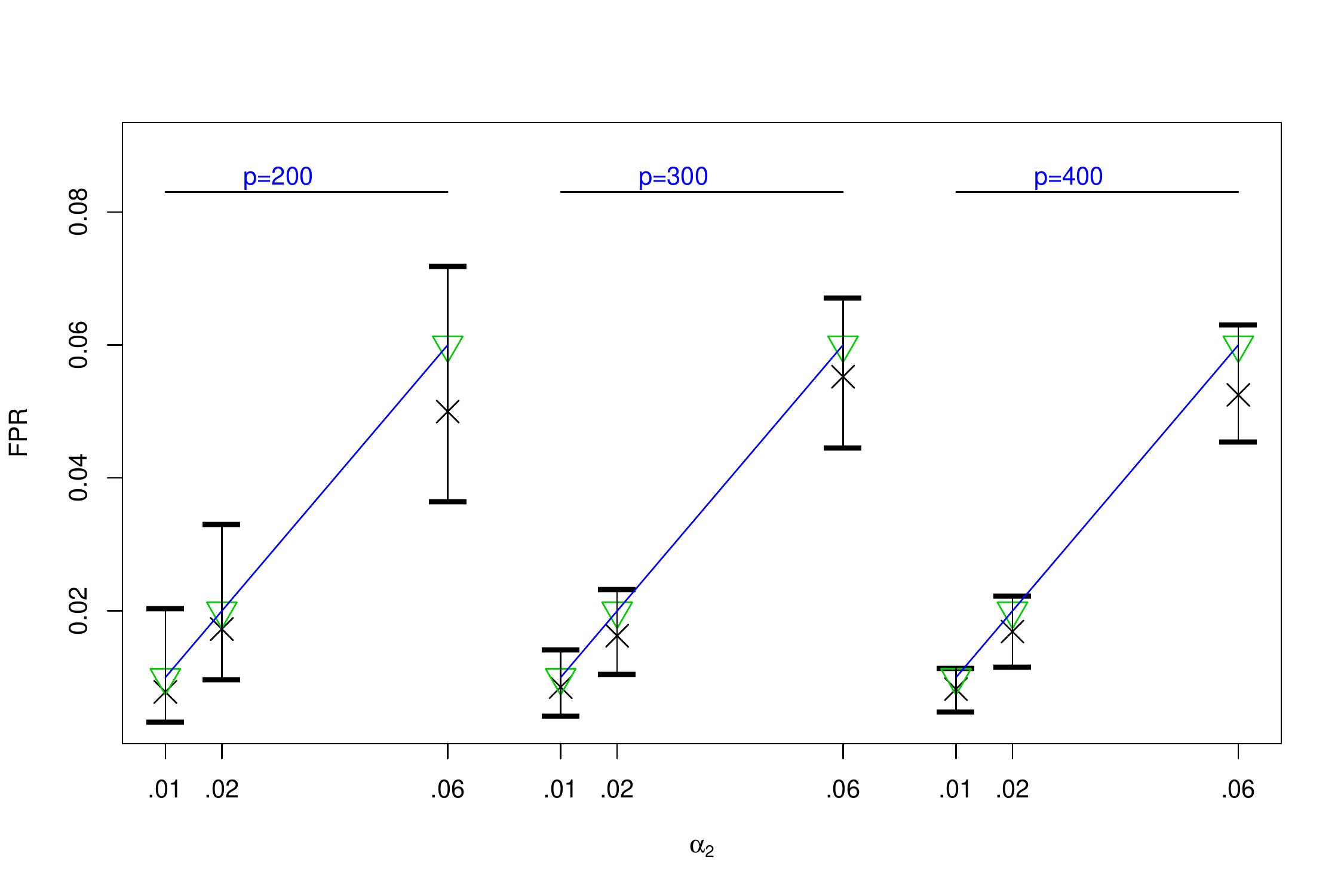}}\\
    \subfloat[n=250]{\includegraphics[scale=0.28]{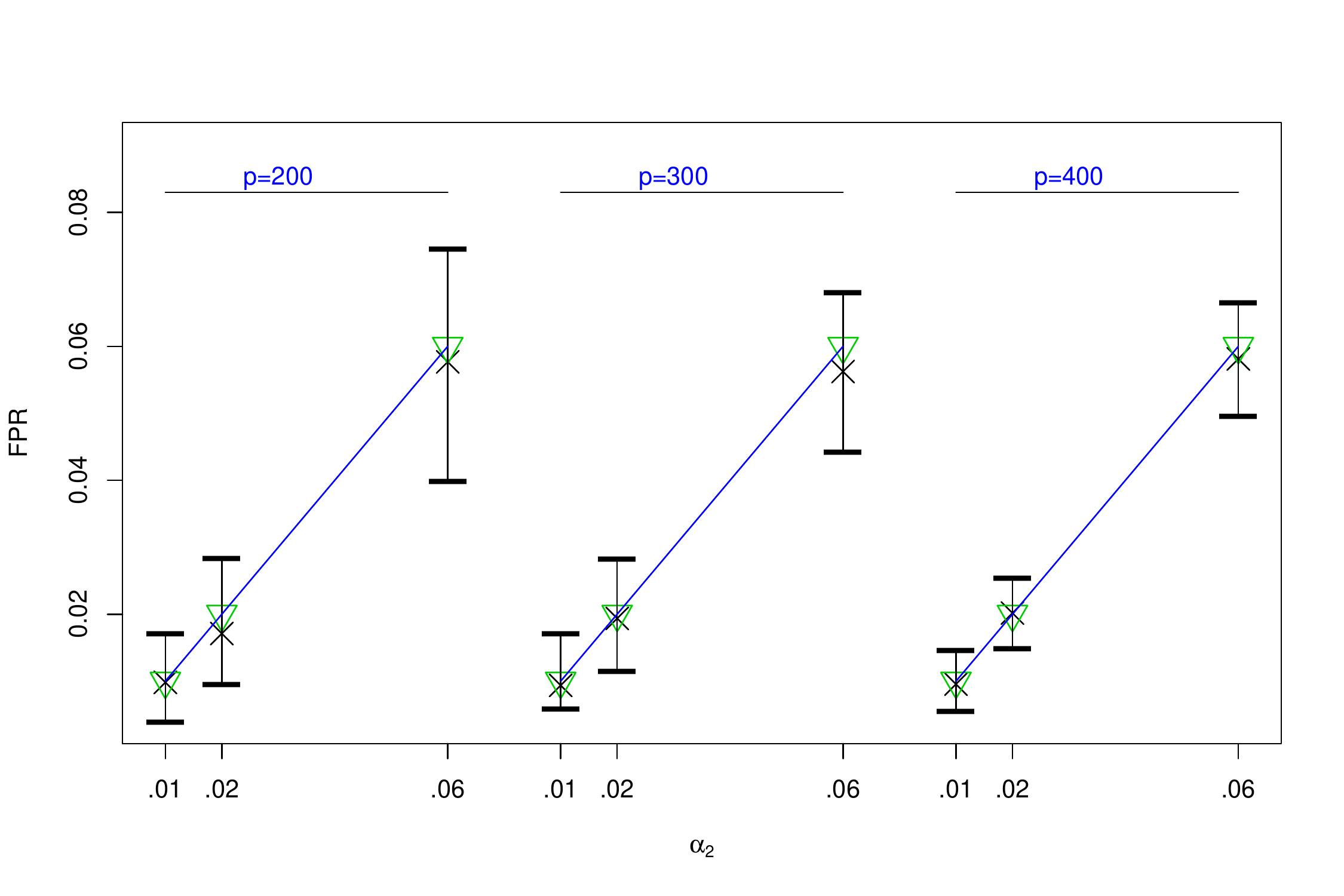}}&
    \subfloat[n=500]{\includegraphics[scale=0.28]{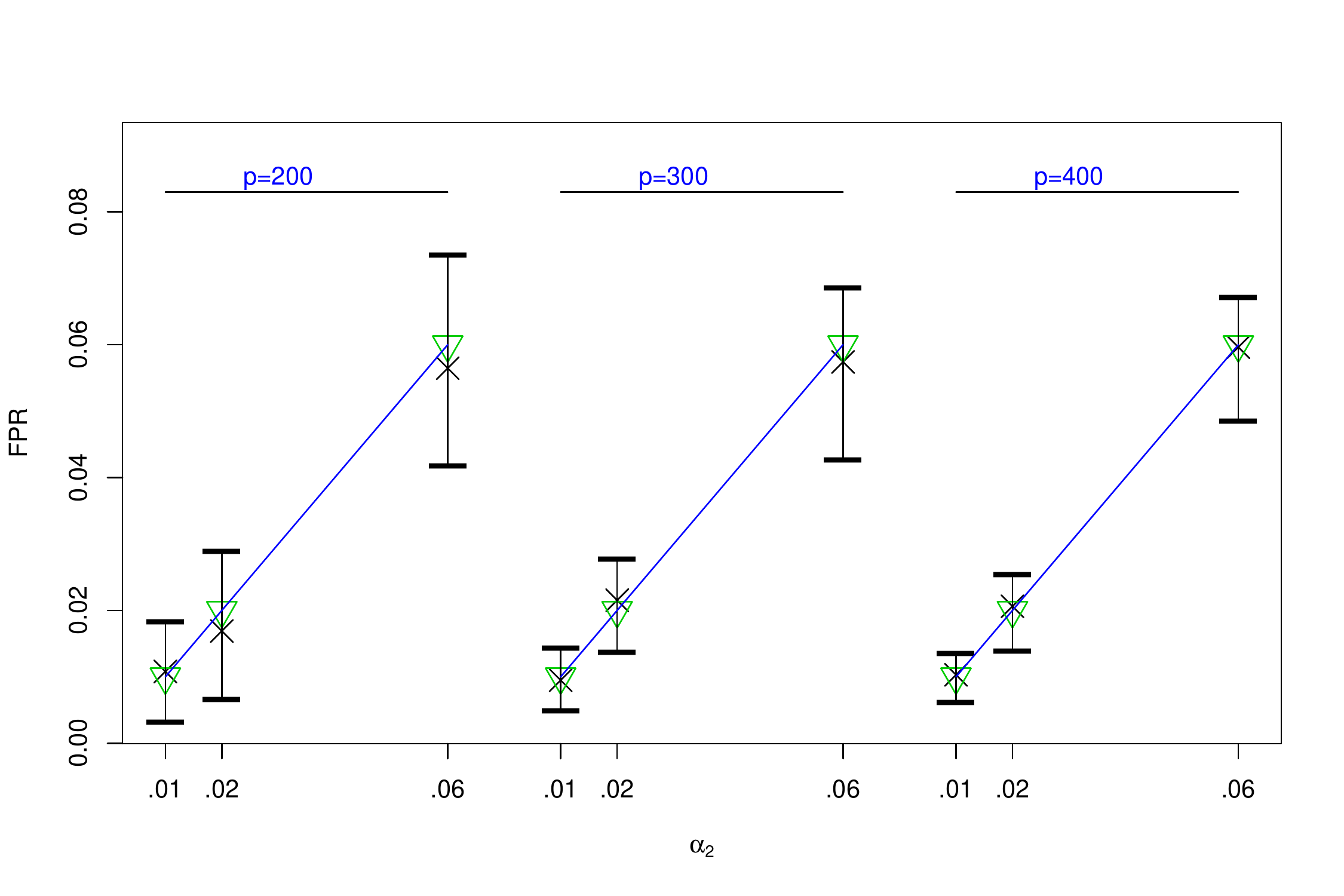}}
 \end{tabular}
 \caption{{\footnotesize FPR vs $\alpha_2$: average (cross) + CI is plotted together with the expected values (triangle). For visualization reasons, x-axis and y-axis are not in the same scale (i.e. $2x : y$)}.}
\label{fSim2C8}
\end{center}
\end{figure}

In Figure \ref{fSim2C8} we compare the expected proportion of false positive edges in the differential network (as defined in Section 3.2 of the article) determined by the value of $\alpha_2$ against the observed false positive rate  (with median and  $95\%$ confidence) using the RCmad estimator to approximate $\sigma_2$ . To draw the confidence interval we replicate the procedure in 100 simulated datasets for different sample sizes and dimension sizes. As for $\alpha_1$, the approximated false positive rate is close to the desired $\alpha_2$ and it is only for very small $n$ that the true value is not included in the confidence interval. 

\footnotesize
\newpage

\bibliography{JointEst240616.bbl}
\end{document}